\documentclass[a4paper,11pt]{article}
\pdfoutput=1 
\RequirePackage{snapshot}  

\usepackage{jheppub} 

\usepackage{bm,amsmath,amssymb,slashed,graphicx,%
            enumerate,alltt,xspace,multirow,xcolor,mathrsfs}
\usepackage{fancyvrb}
\usepackage{booktabs}
\usepackage{graphicx}
\usepackage{subcaption}
\usepackage{xspace}
\usepackage{rotating}
\usepackage[utf8]{inputenc}
\usepackage[export]{adjustbox}
\usepackage{pdflscape}
\usepackage{enumitem}

\makeatletter
\g@addto@macro\bfseries{\boldmath}
\makeatother

\definecolor{labelkey}{rgb}{0,0.5,0.0}

\usepackage{listings}
\definecolor{semiblue}{rgb}{0.3,0.3,0.8}
\newcommand{\logbook}[2]{}

\definecolor{darkgreen}{rgb}{0,0.7,0}
\definecolor{ddarkgreen}{rgb}{0,0.5,0}
\definecolor{grey}{rgb}{0.5,0.5,0.5}
\definecolor{orange}{rgb}{1.0,0.4,0.4}
\definecolor{cyan}{rgb}{0.0,1.0,1.0}
\definecolor{magenta}{rgb}{1.0,0.0,1.0}

\newcommand{\cdiff}{c_{_\delta}}

\newcommand{\GeV}{\;\mathrm{GeV}}
\newcommand{\TeV}{\;\mathrm{TeV}}
\newcommand{\order}[1]{{\cal O}\left(#1\right)}
\newcommand{\as}{\alpha_s}

\newcommand{\abar}{{\bar{\alpha}}}

\newcommand{\rtas}{\sqrt{\alpha_s}}

\newcommand{\qbar}{{\bar q}}
\newcommand{\ktcut}{k_{t,{\text {cut}}}}

\newcommand{\yjet}{y_\text{jet}}
\newcommand{\yhat}{\hat{y}}
\newcommand{\ptmin}{{p_{\perp,{\text{min}}}}}
\newcommand{\nlp}{N^{\text {(Lund)}}}

\newcommand{\avnlp}{\langle N^{\text {(Lund)}}\rangle}
\newcommand{\avnlpbar}{\langle \overline{N^{(\text{Lund})}}\rangle}

\newcommand{\coshnu}{\cosh\nu}
\newcommand{\sinhnu}{\sinh\nu}

\newcommand{\beq}{\begin{eqnarray}}
\newcommand{\eeq}{\end{eqnarray}}

\newcommand{\dd}{\mathrm{d}}
\renewcommand{\l}{{\ell}}

\newcommand{\DL}{\text{DL}\xspace}
\newcommand{\NDL}{\text{NDL}\xspace}
\newcommand{\NNDL}{\text{NNDL}\xspace}

\definecolor{colourndlsqr}{RGB}{255,128,  0} 
\definecolor{coloureloss} {RGB}{255,  0,  0} 
\definecolor{colourclust} {RGB}{255,  0,255} 
\definecolor{colourhme}   {RGB}{  0,128,  0} 
\definecolor{colourpair}  {RGB}{  0,  0,255} 
\definecolor{colourend}   {RGB}{  0,  0,  0} 

\title{Lund multiplicity in QCD jets}

\preprint{CERN-TH-2022-205, OUTP-22-13P}

\newcommand{\OXaff}{Rudolf Peierls Centre for Theoretical Physics, Clarendon 
Laboratory, Parks Road,
  University of Oxford, Oxford OX1 3PU, UK}
\newcommand{\IPhTAff}{Universit\'e Paris-Saclay, CNRS, CEA, Institut de physique 
th\'eorique, 91191, Gif-sur-Yvette, France}
\newcommand{\CERNAff}{Theoretical Physics Department, CERN, 1211 Geneva 23, Switzerland}

\author[a]{Rok Medves,}%
\author[b]{Alba Soto-Ontoso,}%
\author[b,c]{Gregory Soyez}%

\emailAdd{rok.medves@physics.ox.ac.uk}
\emailAdd{alba.soto.ontoso@cern.ch}
\emailAdd{gregory.soyez@ipht.fr}

\affiliation[a]{\OXaff}
\affiliation[b]{\CERNAff}
\affiliation[c]{\IPhTAff}

\date{Received: date / Accepted: \today}

\abstract{
  We compute the average Lund multiplicity of high-energy QCD jets.
  This extends an earlier calculation, done for event-wide multiplicity
  in $e^+e^-$ collisions~\cite{Medves:2022ccw}, to the large energy
  range available at the LHC.
  Our calculation achieves next-to-next-to-double logarithmic (NNDL)
  accuracy.
  Our results are split into a universal collinear piece, common to
  the $e^+e^-$ calculation, and a non-universal large-angle
  contribution.
  The latter amounts to 10-15\% of the total multiplicity.
  We provide accurate LHC predictions by matching our resummed
  calculation to fixed-order NLO results and by incorporating
  non-perturbative corrections via Monte Carlo simulations.
  Including NNDL terms leads to a 50\% reduction of the
  theoretical uncertainty, with non-perturbative corrections
  remaining below $5\%$ down to transverse momentum scales of a few GeV.
  This proves the suitability of Lund multiplicities for robust
  theory-to-data comparisons at the LHC.
}

\begin{document}


\maketitle

\section{Introduction}\label{sec:intro}
Jet substructure observables are instrumental in advancing our
understanding of Quantum Chromodynamics (QCD) at hadron colliders.
Today, infrared-and-collinear (IRC) safe jet substructure
observables are successfully being utilised for a variety of
experimental applications including quark-gluon jet
discrimination~\cite{Gallicchio:2011xq,ATLAS:2014vax,Frye:2017yrw,Dreyer:2020brq},
jet tagging~\cite{Kogler:2018hem,Larkoski:2017jix,Lapsien:2016zor,Krohn:2009zg},
and precision measurements~\cite{Tripathee:2017ybi,Larkoski:2017bvj,ATLAS:2017zda,CMS:2017qlm,CMS:2018ypj,CMS:2018fof,ATLAS:2019mgf,ATLAS:2019kwg,ATLAS:2020bbn,STAR:2020ejj,ALargeIonColliderExperiment:2021mqf,ALICE:2022hyz,ALICE:2022phr}.

Theoretically, the description of jet substructure observables,
especially when accounting for multiple splittings, hinges on
resummation techniques. These techniques typically organise the
perturbative expansion of the observable in terms of single
logarithmic $\as L$ terms, with $L$ a large logarithm of the
considered jet substructure observable. In the literature, the
accuracy of such calculations is regarded as leading logarithmic (LL),
next-to-leading logarithmic (NLL), or generally N$^k$LL. 
NLL accuracy has now become the standard in jet substructure
calculations for proton-proton collisions and several predictions are
available in the literature for both
ungroomed~\cite{Dasgupta:2012hg,Napoletano:2018ohv,Cal:2019hjc,Lifson:2020gua,Reichelt:2021svh,Ziani:2021dxr,Caletti:2021oor,Lee:2022ige,Craft:2022kdo}
and
groomed~\cite{Marzani:2017kqd,Kang:2018jwa,Kang:2019prh,Cal:2019gxa,Anderle:2020mxj,Cal:2020flh,Cal:2021fla}
observables. Quite recently, the first next-to-next-to-leading
logarithmic (NNLL) calculations for the SoftDrop jet mass have been
put forward~\cite{Frye:2016aiz,Kang:2018vgn,Dasgupta:2022fim}. This
increased level of accuracy has opened the path to extractions of $\as$
at hadron colliders, potentially
competitive with fits to lepton collider data~\cite{Hannesdottir:2022rsl}.

However, some observables, including the one studied
in this paper, do not exponentiate and hence do not follow the single-logarithmic
(N$^k$LL) resummation structure.
Instead one relies on an expansion starting from double-logarithms
(\DL) and resumming terms proportional to $\as^n L^{2n-k}$ at N$^k$DL
accuracy.
With respect to this, calculations related to subjet multiplicity have
achieved next-to-double logarithmic (\NDL)
accuracy~\cite{Gerwick:2012fw,Bhattacherjee:2015psa,Medves:2022ccw}.
This approach has also proven useful for resumming advanced substructure
techniques like dynamical grooming observables~\cite{Caucal:2021bae}
whose resummation structure does not exponentiate, in particular for
Sudakov-safe~\cite{Larkoski:2015lea} observables. These calculations
are known to next-to-next-to-double logarithmic (\NNDL) accuracy.
These first-principle calculations of jet substructure observables can
then be directly compared to both experimental
data~\cite{Tripathee:2017ybi,Larkoski:2017bvj,ATLAS:2017zda,CMS:2017qlm,CMS:2018ypj,CMS:2018fof,ATLAS:2019mgf,ATLAS:2019kwg,ATLAS:2020bbn,STAR:2020ejj,ALargeIonColliderExperiment:2021mqf,ALICE:2022hyz,ALICE:2022phr}
and/or full-fledged Monte Carlo simulations.\footnote{Note that in
  some phase-space regions, non-perturbative effects such as
  hadronisation or multi-parton interactions cannot be neglected and
  the perturbative results have to be complemented with a
  non-perturbative factor, determined by either Monte Carlo or
  phenomenological modelling.}

In recent years, the so-called Lund-plane
techniques~\cite{Dreyer:2018nbf} have proven to be a powerful tool for
addressing a wide range of jet substructure
questions~\cite{ATLAS:2020bbn,Lifson:2020gua,Dreyer:2020brq,Dreyer:2021hhr,Medves:2022ccw}.
The most recent observable in this framework has been the {\it Lund multiplicity} 
recently introduced in Ref.~\cite{Medves:2022ccw}. This new observable counts
the number of jets in a hemisphere above a certain momentum cutoff
$\ktcut$.
In general, this number fluctuates from event to event. We focus here
on its average which can be more easily brought to higher-order
calculations in perturbation theory.
The presence of two disparate scales, i.e.\ the hard scale $Q$ and the infrared cutoff, leads to the appearance 
of large logarithms $L\equiv \ln(Q/\ktcut)$ that must be resummed to all-orders in order to guarantee
the convergence of the perturbative expansion. The resummation structure of this observable can 
be organised in terms of double logarithms as follows:
\begin{equation}
\label{eq:log-counting}
  \langle N^{\text{(Lund)}}(\as; L)\rangle
  =  \bigg[
    \underbrace{h_1(\as L^2)}_{\DL}
    + \underbrace{\rtas h_2(\as L^2)}_{\NDL}
    + \underbrace{\as h_3(\as L^2)}_{\NNDL} + \dots
  \bigg]
  + \mathcal{O}\big(e^{-|L|}\big)\, ,
\end{equation}
with $L\equiv \ln(Q/\ktcut)$. In Ref.~\cite{Medves:2022ccw},
$\avnlp$ was calculated at \NNDL accuracy for $e^+e^-$ collisions, i.e.\ up to and including
$h_3$ in Eq.~\eqref{eq:log-counting}, and at \NDL accuracy 
for colour singlet production at hadron colliders.

In the present manuscript, we extend the calculation to jets produced
in high-energy proton-proton collisions, achieving \NNDL accuracy. To
that end, we first generalise the Lund-multiplicity definition so as
to count the number of subjets inside high-energy jets. As any other
Lund-plane-based observable, the procedure begins with reclustering a
jet of interest at a high-energy collider with the Cambridge-Aachen
algorithm~\cite{Dokshitzer:1997in, Wobisch:1998wt}.\footnote{
  We note that using a different recombination scheme for the
  Cambridge-Aachen reclustering procedure, such as the
  winner-takes-all~\cite{Larkoski:2014uqa} recombination scheme, only
  brings differences beyond our target \NNDL accuracy (cf.\
  appendix~A.2 of Ref.~\cite{Medves:2022ccw}).}
Then, its Lund multiplicity is defined as follows:
\newpage
\begin{enumerate}[noitemsep,topsep=2pt,parsep=2pt]
\item Set $\nlp=1$.
\item \label{item:lund-multiplicity-step2}
  Undo the last clustering step to get two subjets $j_1$ and
  $j_2$, with $j_1$ the harder branch, i.e.\ $p_{t1}>p_{t2}$.
\item \label{item:lund-multiplicity-step3}
Calculate the relative transverse momentum of the splitting as
\begin{align}
  k_t \equiv \min(p_{t1}, p_{t2}) \Delta_{12}
  \qquad \text{where} \qquad
  \Delta_{12}^2 = (y_1 - y_2)^2 + (\phi_1 - \phi_2)^2,
\label{kt:def}
\end{align}
  with $y_{1,2}$ ($\phi_{1,2}$) the rapidities (azimuthal
  angles) of the subjets with respect to the beam.

\item \label{item:lund-multiplicity-step4}
  If $k_t\ge\ktcut$ the splitting contributes to the Lund
  multiplicity, i.e.\ $\nlp$ is incremented by one, and we go back to
  step~\ref{item:lund-multiplicity-step2} for {\it each} of the two subjets.
\item Otherwise, if $k_t<\ktcut$, repeat from step 
\ref{item:lund-multiplicity-step2} following only the harder subjet $j_1$.
\end{enumerate}
The procedure terminates when there is nothing left to
decluster.
The result of the above algorithm is one plus the total number of
emissions in both the primary Lund plane and subsidiary Lund leaves
whose transverse momentum is above $\ktcut$. Naturally, the average
Lund multiplicity in a high-energy jet follows the resummation
structure of Eq.~\eqref{eq:log-counting}.
The main complication compared to the earlier calculation of the Lund
multiplicity in $e^+e^-$ events comes from a richer structure of
large-angle gluon radiation.
In this context, the treatment of gluon radiation at large angles is
known to be prohibitively complex if one wants to retain full-colour
accuracy.  
In this work, we will target full-colour accuracy for the first
large-angle emission but limit ourselves to a leading-colour
prescription for subsequent large-angle emissions. 

The paper is organised as follows. In section~\ref{sec:recape+e-} we present a brief overview
of the \NNDL calculation of $\avnlp$ in $e^+e^-$ collisions that was
extensively discussed in Ref.~\cite{Medves:2022ccw}. Then, we move on to 
the novel ingredients of the $pp$ calculation in
section~\ref{sec:pp-calc} which start appearing at \NDL accuracy.
We supplement our resummed calculation with a matching to the exact
$\order{\as^2}$ result and a study of its sensitivity to
non-perturbative corrections in section~\ref{sec:numerics}. The final
phenomenological predictions for LHC energies can be found in
section~\ref{sec:pheno}.  We conclude with a summary of our main
findings in section~\ref{sec:conclusions}.

\section{Recap of average Lund multiplicity in $e^+e^-$ collisions}
\label{sec:recape+e-}
The purpose of this section is to summarise the 
working principles of the resummation strategy that
we pursue for Lund multiplicity. An extended discussion and
full details can be found in Ref.~\cite{Medves:2022ccw}.
We begin by considering the Lund multiplicity of a 
$q/g$-initiated hemisphere in an $e^+e^-$ collision at
a centre-of-mass energy $Q$.
Note that in this case, we use energies and angles instead of
$p_\perp$ and $\Delta$, with the relative transverse momentum defined
as $k_t=\min(E_1,E_2)\sin \theta_{12}$, which matches the $pp$
definition in the collinear limit.
Our aim is to compute the average number of clusterings,
within this hemisphere, that satisfy the $k_t >\ktcut$ condition at \NNDL accuracy.
To this end, we calculate the $h_1,h_2$ and $h_3$ functions entering 
into Eq.~\eqref{eq:log-counting}. 

\paragraph{\DL accuracy.}
We begin by discussing the case in which a parton 
with energy $Q$ radiates a set of $n$ soft-and-collinear gluons,
that is, the double-logarithmic (\DL) limit.
In this case, emissions are strongly ordered both in
energy, i.e.\ $Q\gg E_{1} \gg E_{2} \gg \ldots \gg E_{n}$, 
and in angle.  Further, we can take the coupling constant to be fixed.
The first important observation is that, at
\DL accuracy, the only non-zero contribution comes from the kinematic
configuration where the second gluon is emitted from the first gluon
(i.e.\ with $\theta_{21}\ll \theta_{10}$ in the collinear limit), the
third gluon is emitted from the second ($\theta_{32}\ll \theta_{21}$),
etc...
All the other configurations cancel between real and virtual
contributions.
In terms of colour factors, it means that the only non-zero
contribution comes with a colour factor $C_iC_A^{n-1}$ with $C_i=C_F$
($C_i=C_A$) for a quark-initiated (gluon-initiated) jet.
In what follows we will refer to this configuration as a nested chain of
double-logarithmic emissions.

To compute the \DL Lund multiplicity, we first write the angular ordering
condition as $0 < \eta_1\ll \eta_2 \ll \ldots \ll \eta_n$, where
$\eta_j = -\ln\tan(\theta_{j, j-1} / 2)$, $\theta_{j,j-1}$ is the
angle between the $j^\text{th}$ and $(j-1)^\text{th}$ emission and
$\theta_{1,0}\equiv \theta_1$ is the angle between the first emission
and the initial hard parton.
At \DL accuracy, the differential probability for each emission can be
written as
$\abar \frac{\dd k_t}{k_t} \dd\eta=\abar \frac{\dd x}{x} \dd\eta$,
with $\abar = 2\as C_A/\pi$, $x$ the energy fraction of the emission
and $k_t$ its relative transverse momentum with respect to the
emitter.
By iterating this branching probability for $n$ emissions, and
imposing angular and energy ordering, we obtain the Lund
multiplicity at \DL accuracy
\begin{align}\label{eq:final-master-dl}
 N_i^{(\text{DL})} \equiv h^{(i)}_{1, e^+e^-}
  & = 1+ \frac{C_i}{C_A} \sum_{n=1}^\infty  \abar^n
    \int_{0}^\infty\dd\eta_1\int_{\eta_1}^{\infty}\dd\eta_2 \dots
    \int_{\eta_{n-1}}^\infty \dd\eta_n \nonumber\\
  & \phantom{= 1 + \frac{C_i}{C_A} \sum_{n=0}^\infty \abar^n}
    \int_0^1\frac{\dd x_1}{x_1}\int_0^{x_1}\frac{\dd x_2}{x_2} \dots
    \int_0^{x_{n-1}} \frac{\dd x_n}{x_n} \Theta(x_n 
    e^{-\eta_n}>e^{-L})\nonumber\\ 
  &= 1+\frac{C_i}{C_A}\left(\coshnu-1\right),
    \quad
    \text{with }
    \nu = \sqrt{\abar L^2}, \abar = \frac{2\as C_A}{\pi}.
\end{align}
In this expression, the `1' accounts for the jet initiator. Similarly,
the differential distribution of Lund declusterings at a given $k_t$
is given by
\begin{equation}
\label{eq:nDL}
n_i^{(\DL)} = \frac{\dd N_i^{(\text{DL})} }{\dd L} =\frac{C_i}{C_A}\sqrt\abar \sinh \nu.
\end{equation}
\begin{figure}
  \begin{subfigure}[t]{0.49\linewidth}
  \centering
    \includegraphics[scale=1]{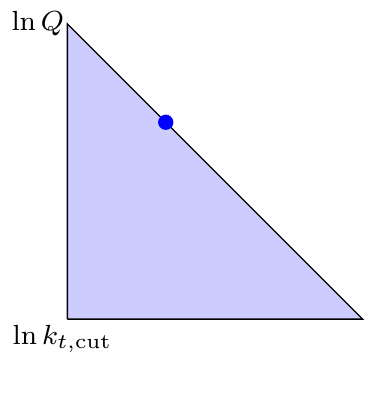}
    \caption{Hard collinear splitting}
    \label{fig:ndl-diagram-hc}
  \end{subfigure}
   \begin{subfigure}[t]{0.49\linewidth}
   \centering
    \includegraphics[scale=1]{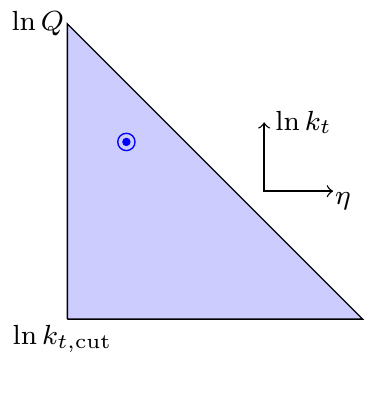}
    \caption{Running coupling}
    \label{fig:ndl-diagram-b0}
  \end{subfigure}
  \caption{Lund representation of the \NDL configurations that are
    identical in $e^+e^-$ collisions and $pp$ jets.}\label{fig:ndl-ee}
\end{figure}

\paragraph{\NDL accuracy.} A key aspect of our resummation strategy is
that in order to achieve \NDL accuracy one only has to lift the
soft-and-collinear constraint for a single emission in the nested
chain of double-logarithmic emissions. This
\NDL-like emission occurs at a scale $\ell = \ln(Q/k_t)$ and can
be either real or virtual and either primary or secondary.
Let us denote by $K_\NDL$ the genuine \NDL corrections to the matrix
element for this emission.  All other emissions preceding and
following this \NDL emission down to the infrared cutoff $\ktcut$ are
then treated in the \DL, soft-and-collinear, approximation and
resummed using Eq.~\eqref{eq:final-master-dl} or \eqref{eq:nDL}.
The master formula that embodies the above ideas and gives the pure
\NDL correction to the Lund multiplicity reads
\begin{align}
  \delta N^{(\NDL)}_i
  &= \int_0^L  \dd \l \,\Big\{
    K^R_\NDL\,
     \big[N^{(\DL)}_\text{hard}(L;\l)+N^{(\DL)}_\text{soft}(L;\l)\big]
    - K^V_\NDL N^{(\DL)}_i(L;\l)
    \Big\}  
         \label{eq:n-ndl-full}\\
   &+ \int_0^L\dd \l_1 
     n^{(\DL)}_i(\l_1) \int_{\l_1}^L  \dd \l_2\,
     \Big\{K^R_\NDL
     \big[N^{(\DL)}_\text{hard}(L;\l_2)+N^{(\DL)}_\text{soft}(L;\l_2)\big]
     - K^V_\NDL
     N^{(\DL)}_g(L;\l_2)\Big\},\nonumber
\end{align}
where $N^{(\DL)}(L; \ell) \equiv N^{(\DL)}(L-\ell)$ was defined in
Eq.~\eqref{eq:final-master-dl}.
The first line of Eq.~\eqref{eq:n-ndl-full} accounts for the case in which the \NDL emission
takes place along the primary branch, while the second line describes
a chain of \DL emissions between scales $Q$ and $Qe^{-\ell_1}$, followed by
an \NDL correction in some secondary branch between scales $Qe^{-\ell_1}$ and
$Qe^{-\ell_2}$, followed again by \DL emissions between scales
$Qe^{-\ell_2}$ and $\ktcut = Qe^{-L}$.
For corrections associated with a real emission (terms involving
$K_{\NDL}^{R}$) one includes subsequent branchings on both the hard
and soft branches, as indicated by the subscripts in the squared
brackets of Eq.~(\ref{eq:n-ndl-full}).

In order to systematically identify all \NDL corrections we first
split them into two categories: corrections related to the running
of the strong coupling and kinematic corrections. The former
corresponds to 1-loop running coupling 
corrections, i.e.\ $\as \to \as(k_t)\simeq\as (1+ 2\as\beta_0\ln Q/k_t)$.  A
powerful tool for listing all corrections of kinematic origin is the Lund
plane representation~\cite{Dreyer:2018nbf}, in which the location of
a given emission in the Lund plane reveals at which accuracy it
contributes.
For example, a \DL emission lives in the bulk of the Lund plane, while
\NDL ones populate regions of the Lund plane where the integrated
matrix element is single-logarithmic, i.e.\ scales as $\as L$. This is
the case when the emission is either hard-collinear, wide-angle or has
$k_t\sim\ktcut$. Of these, Ref.~\cite{Medves:2022ccw} showed that only
the hard-collinear yields a \NDL correction.
The \NDL Lund multiplicity, including the running-coupling and
hard-collinear corrections, displayed in figure~\ref{fig:ndl-ee}, is
found to be
\begin{subequations}
  \label{eq:h2-coll}
  \begin{align}
    h_{2,e^+e^-}^{(q)}  =
    h_{2,\text{coll}}^{(q)}
    & = \frac{C_F}{\sqrt{2\pi C_A}}
     \bigg\{\frac{\pi\beta_0}{2C_A}
      \big[(\nu^2-1)\sinhnu + \nu\coshnu \big] + (B_{gg}+ \cdiff B_{gq} ) \nu\coshnu\\
    & \phantom{ = \frac{C_F}{\sqrt{2\pi C_A}}\;}
      + \big[2B_q- B_{gg} + \left(2-3\cdiff\right)B_{gq}\big]\sinhnu
      + 2(\cdiff-1)B_{gq} \nu\bigg\},\nonumber\\
    h_{2,e^+e^-}^{(g)}  =
    h_{2,\text{coll}}^{(g)}
    & = \sqrt{\frac{C_A}{2\pi}}
    \bigg\{ \frac{\pi\beta_0}{2C_A}
      \big[(\nu^2-1)\sinhnu + \nu\coshnu \big] \\
    & \phantom{ = \sqrt{\frac{C_A}{2\pi}}\;}
      + (B_{gg}+ \cdiff B_{gq}) \nu\coshnu 
      + \big[B_{gg} + (2-\cdiff)B_{gq}\big] \sinhnu\bigg\},\nonumber
  \end{align}
\end{subequations}
for quark and gluon hemispheres respectively. Here,
$\cdiff = (2C_F - C_A)/C_A$, and the values of the hard-collinear 
$B_{ij}$ coefficients can be found in
table~\ref{table:multiplicity-coefficients-ee}.
The subscript ``coll'' highlights that these corrections are
purely in the collinear region, which will be useful later.
%

\paragraph{\NNDL accuracy.}
The same logic applies when going one order higher in logarithmic
accuracy, where only a finite number of emissions should be treated
beyond the soft-and-collinear approximation.
Again, it is useful to distinguish kinematic corrections from running coupling effects.
Let us first focus on the former.
An \NNDL contribution of the form $\as^n L^{2n-2}$ can be obtained
either through one genuine ``\NNDL-like'' emission contributing a
factor $\as$ together with $(n-1)$ \DL, soft-and-collinear, emissions,
i.e.\ $\as^n L^{2n-2}=(\as)(\as L^2)^{n-1}$, or through two \NDL-like
emissions dressed with $(n-2)$ \DL emissions, i.e.\
$\as^n L^{2n-2}=(\as L)^2(\as L^2)^{n-2}$.
\NNDL running coupling corrections  arise
through terms proportional to
$\beta_0^2$, generated by either one or two emissions, and mixed cases in which one emission carried a 
$\beta_0$-term and another pertains to the hard-collinear regime. 
Note that two-loop running-coupling corrections, proportional to
$\beta_1$, would only start appearing at N$^3$DL.

Hence, in order to resum the average Lund multiplicity to \NNDL accuracy, one
must: (i) systematically identify the relevant contributions, (ii) for
\NNDL-like contributions, compute the corresponding kernel $K_\NNDL$
by means of a fixed-order calculation, and (iii) perform the
resummation using Eq.~\eqref{eq:n-ndl-full} with $K_\NDL$ replaced by
$K_\NNDL$, or use a similar equation with two \NDL insertions for
contributions of the form $(\as L)^2(\as L^2)^{n-2}$.
\begin{figure}
  \begin{subfigure}[t]{0.24\linewidth}
  \centering
    \includegraphics[scale=1]{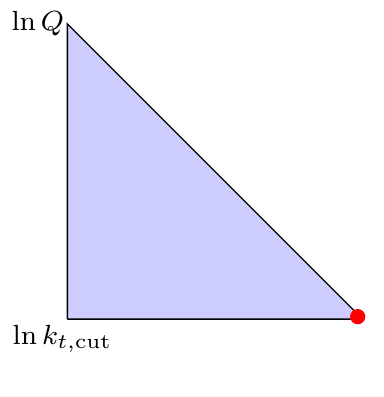}
    \caption{Collinear endpoint.}
    \label{fig:nndl-diagram-hccut}
  \end{subfigure}
   \begin{subfigure}[t]{0.24\linewidth}
   \centering
    \includegraphics[scale=1]{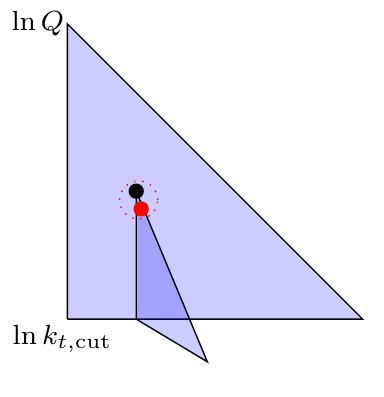}
    \caption{Close-by pair.}
    \label{fig:nndl-diagram-db}
  \end{subfigure}
     \begin{subfigure}[t]{0.24\linewidth}
   \centering
    \includegraphics[scale=1]{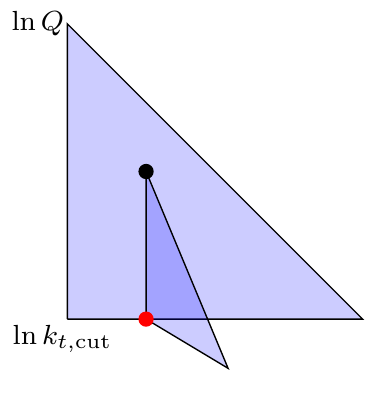}
    \caption{Clustering.}
    \label{fig:nndl-diagram-clust}
  \end{subfigure}
  \begin{subfigure}[t]{0.24\linewidth}
  \centering
    \includegraphics[scale=1]{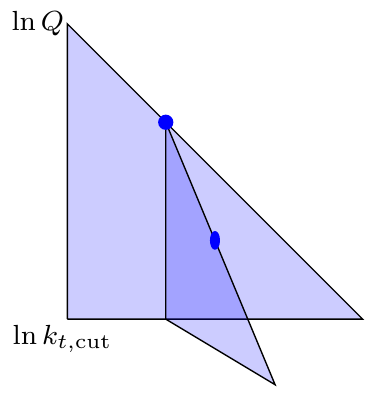}
    \caption{Two hard-collinear.}
    \label{fig:nndl-diagram-hcxhc}
  \end{subfigure}\\[3mm]
   \begin{subfigure}[t]{0.49\linewidth}
   \centering
    \includegraphics[width=0.485\textwidth]{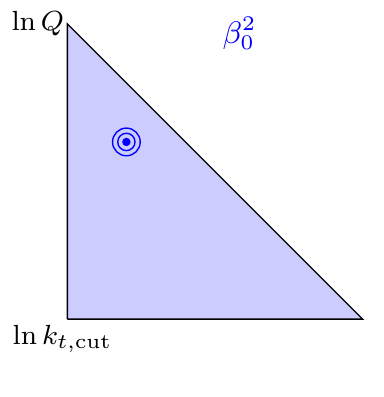}
    \includegraphics[width=0.485\textwidth]{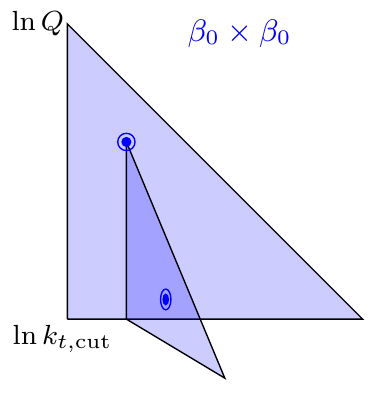}
    \caption{Squared running coupling.}
    \label{fig:nndl-diagram-b0xb0}
  \end{subfigure}
      \begin{subfigure}[t]{0.49\linewidth}
   \centering
    \includegraphics[width=0.485\textwidth]{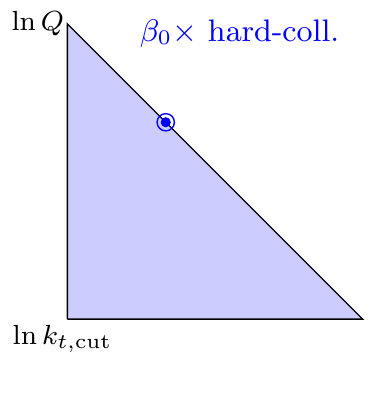}
    \includegraphics[width=0.485\textwidth]{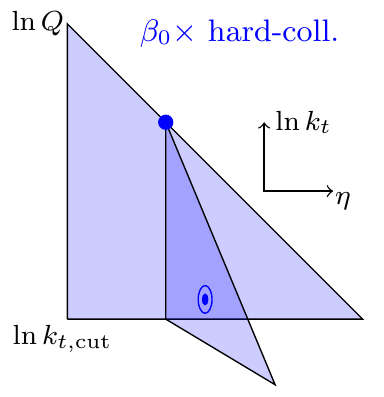}
    \caption{Running coupling $\times$ hard-collinear.}
    \label{fig:nndl-diagram-b0xhc}
  \end{subfigure}
  \caption{Lund representation of the \NNDL configurations that are
    identical in $e^+e^-$ and $pp$ collisions. Black dots indicate
    emissions which contribute a factor $\as L^2$, blue dots emissions
    which contribute a factor $\as L$, and red dots emissions which
    contribute as $\as$. Running coupling corrections are represented
    by open circles.} \label{fig:nndl-ee}
\end{figure}
In total, 8 non-zero contributions contribute to the average Lund
multiplicity at \NNDL accuracy.
The result can be written as
\begin{align}\label{eq:h3-ee}
  h^{(q,g)}_{3,e^+e^-} = h^{(q,g)}_{3,\text{coll}} + h^{(q,g)}_{3,e^+e^-,\text{la}},
\end{align}
with
\begin{align}
  2\pi h^{(q)}_{3,e^+e^-,\text{la}}
  & = D_\text{hme}^{qqg}\coshnu,\label{eq:h3-la-ee-q}\\
  \label{eq:h3-coll-q}
  2\pi h^{(q)}_{3,\text{coll}}
  & = D_{\text{end}}^{q\to qg}+ 
    \left(D_{\text{end}}^{g\to gg}+D_{\text{end}}^{g\to q\bar q}\right)
    \frac{C_F}{C_A}(\coshnu-1) \\
  & 
    + {\frac{C_F}{C_A}
    \Big[
    (1-\cdiff) D_{\text{pair}}^{q\qbar}
    (\coshnu-1)
    +\left(K + D_{\text{pair}}^{gg} + \cdiff D_{\text{pair}}^{q\qbar}\right)
    \frac{\nu}{2}\sinhnu
    \Big]}\nonumber\\
  & + {C_F
    \Big[\Big(
    \coshnu-1 -\frac{1-\cdiff}{4}\nu^2\Big)
    D_\text{clust}^\text{(prim)} + (\coshnu-1)
    D_\text{clust}^\text{(sec)}\Big]}\nonumber\\
  & 
  + {\frac{C_F}{C_A}
    \Big[
    D_{\text{e-loss}}^g \, \frac{\nu}{2} \sinhnu
    + \left(D_{\text{e-loss}}^q  -
    D_{\text{e-loss}}^g\right)(\coshnu - 1) \Big]}
    \nonumber\\
  & + {\frac{C_F}{2}
    \Big\{
    (B_{gg}+\cdiff B_{gq})^2\nu^2 \coshnu
    +8\left[2\cdiff B_{gg}- 2\cdiff B_q
    -(1-3\cdiff^2)B_{gq}\right]B_{gq}\coshnu}\nonumber\\
  & \phantom{+C_F}{
    +\left[4 B_q(B_{gg}+(2 \cdiff+1)B_{gq})-(B_{gg}+\cdiff
    B_{gq})(B_{gg}+9\cdiff B_{gq})\right]\nu\sinhnu
    }\nonumber\\
  & \phantom{+C_F}{
    +4(1-\cdiff^2)B_{gq}^2\nu^2+8\left[2\cdiff B_q
    -2\cdiff B_{gg}+(1-3\cdiff^2)B_{gq}\right]B_{gq}
    \Big\}}\nonumber\\
  & +{
    \frac{C_F}{C_A} \frac{\pi\beta_0}{2} \Big\{
    (B_{gg}+\cdiff B_{gq})\nu^3\sinhnu
    +\left[2B_q-2B_{gg}+(6-8\cdiff)B_{gq}\right]\nu\sinhnu}
    \nonumber\\
  & \phantom{+ C_F\pi\beta_0}
    {
    +2 (B_q+B_{gg}+B_{gq})\nu^2\coshnu
    -4(1-\cdiff)B_{gq}(2\coshnu-2+\nu^2)\Big\}}
    \nonumber\\    
  & + {\frac{C_F}{C_A}
    \frac{\pi^2 \beta_0^2}{8C_A}
    \big[
    3\nu(2\nu^2-1) \sinhnu+(\nu^4+3\nu^2) \coshnu  \big]},
    \nonumber
\end{align}
for quark-initiated jets, and
\begin{align}
  2\pi h^{(g)}_{3,e^+e^-,\text{la}}
  & = D_\text{hme}^{ggg}\coshnu
    + D_\text{hme}^{gq\bar q}(\cdiff\coshnu+1-\cdiff),\label{eq:h3-la-ee-g}\\
  \label{eq:h3-coll-g}
  2\pi h^{(g)}_{3,\text{coll}}
  & = 
    \left(
    D_{\text{end}}^{g\to gg} + D_{\text{end}}^{g\to q\qbar}
    \right) \coshnu \\
  & + {
    \left[
    (1-\cdiff) D_{\text{pair}}^{q\qbar} (\coshnu-1)
    +\left(K + D_{\text{pair}}^{gg} + \cdiff D_{\text{pair}}^{q\qbar}\right)
    \frac{\nu}{2}\sinhnu
    \right]} \nonumber\\
  & + {C_A
    \left(D_\text{clust}^\text{(prim)}+ D_\text{clust}^\text{(sec)}\right)
    (\coshnu-1)}
    + {
    D_{\text{e-loss}}^g \, \frac{\nu}{2} \sinhnu}\nonumber\\
  & + {
    \frac{C_A}{2}
    \big\{
    (B_{gg}+\cdiff B_{gq})^2\nu^2 \coshnu
    -8(1-\cdiff^2)B_{gq}^2(\coshnu-1)}\nonumber\\
  & \phantom{+C_A}{
    +\left[(B_{gg}+\cdiff B_{gq})(3B_{gg}-5\cdiff
    B_{gq})+4(1+\cdiff)B_{gq}B_q\right]\nu\sinhnu\big\}}\nonumber\\
  & + {
    \frac{\pi\beta_0}{2} \big\{
      (B_{gg}+\cdiff B_{gq})\nu^3\sinhnu +6(1-\cdiff)B_{gq} \nu\sinhnu
    +2\left[2B_{gg}+(1+\cdiff)B_{gq}\right]\nu^2\coshnu}
    \nonumber\\    
  & \phantom{+\pi\beta_0}{
    -8B_{gq}(1-\cdiff)(\coshnu-1)\big\}}
    + {\frac{\pi^2 \beta_0^2}{8C_A}
    \left[
    3\nu(2\nu^2-1) \sinhnu + (\nu^4+3\nu^2) \coshnu
    \right]},\nonumber
\end{align}
for gluon-initiated jets.

\begin{table}
  \renewcommand{\arraystretch}{1.3}
  
  \begin{center}
    \begin{tabular}{l c}
      \toprule
      Contribution
      & Coefficient\\
      \midrule
      running coupling
      & $\beta_0=\frac{11C_A - 4n_fT_R}{12\pi}$ \\
      hard-collinear correction
      & $B_q=-\frac{3}{4}$, $B_{gg}=-\frac{11}{12}$, $B_{gq}=\frac{n_fT_R}{3C_A}$ \\
      \midrule
      hard matrix-element ($e^+e^-$)
      & $D_{\text{hme}}^{qqg}=C_F(\frac{\pi^2}{6}-\frac{7}{4})$,\\
      & $D_{\text{hme}}^{ggg}=C_A(\frac{\pi^2}{6}-\frac{49}{36})$,
        $D_{\text{hme}}^{gq\bar q}=n_fT_R\frac{2}{9}$\\
      collinear endpoint
      & $D_{\text{end}}^{q \to qg}    = C_F(3+3\ln 2-\frac{\pi^2}{3})$ \\
      & $D_{\text{end}}^{g \to gg}    = C_A(\frac{137}{36}+\frac{11}{3}\ln 2-\frac{\pi^2}{3})$ \\
      & $D_{\text{end}}^{g \to q\qbar} = n_fT_R(-\frac{29}{18}-\frac{4}{3}\ln 2)$ \\
      commensurate $k_t$ and angle
      & $D_{\text{pair}}^{q\qbar}=\frac{13}{9}n_fT_R$, $D_{\text{pair}}^{gg}=(\frac{\pi^2}{6}-\frac{67}{18})C_A$\\
      & $K = \left(\frac{67}{18}-\frac{\pi^2}{6} \right)C_A - \frac{10}{9} n_f T_R$\\
      clustering
      & $D_\text{clust}^\text{(prim)}=-\frac{5\pi^2}{54}$,
        $D_\text{clust}^\text{(sec)}=\frac{\pi^2}{27}$\\
      energy loss
      & $D_{\text{e-loss}}^q=\frac{7}{2}C_A+(\frac{5}{2}-\frac{2\pi^2}{3})C_F$\\
      & $D_{\text{e-loss}}^g=(\frac{67}{9}-\frac{2\pi^2}{3})C_A-\frac{26}{9}\frac{C_F}{C_A}n_fT_R$\\
      \bottomrule
    \end{tabular}
  \end{center}
  \caption{Coefficients entering the \NDL $h_{2,e^+e^-}$ (top entries)
    and \NNDL $h_{3,e^+e^-}$ functions (bottom entries) for the
    average Lund multiplicity in $e^+e^-$ events.}
  \label{table:multiplicity-coefficients-ee}
\end{table}

The values of the coefficients entering in Eq.~\eqref{eq:h3-ee} 
can be found in table~\ref{table:multiplicity-coefficients-ee}.
Note that the results in the above expressions have been explicitly
separated in two parts: (i) $h_{3,\text{coll}}^{(q,g)}$ that involves
the \NNDL contributions sensitive to collinear physics 
and running-coupling corrections, see figure~\ref{fig:nndl-ee},
and (ii) $h_{3,\text{la}}^{(q,g)}$ that includes all terms
involving large-angle kinematics.
As we will argue in the next section, the collinear contributions can
be directly recycled for the case of the Lund multiplicity in a
high-energy jet, while the large-angle contributions have to be
recomputed.

\section{Calculation of the average Lund multiplicity in jets}
\label{sec:pp-calc}

In this section, we derive the main result of this paper, namely we
extend the calculation of the Lund multiplicity to the case of the
subjet multiplicity within a high-energy jet.

We therefore start with a jet of transverse momentum $p_\perp$, and
rapidity $y_\text{jet}$, assuming that it has been reconstructed using the
anti-$k_t$ jet algorithm~\cite{Cacciari:2008gp} with a jet radius $R$.
The average Lund multiplicity
$\langle N^{\text{(Lund)}}_\text{jet}(\as;L)\rangle$ above a relative
transverse momentum cut, $\ktcut$, follows the logarithmic expansion
of Eq.~(\ref{eq:log-counting}) with the logarithm $L$ defined as
$L\equiv L_\text{jet}=\ln(p_\perp R/\ktcut)$.
In practical applications, the transverse momentum and rapidity of the
jet will not be fixed but integrated over a fiducial phase space. We
discuss how to translate our results for a fixed jet $p_\perp$ and
$y_\text{jet}$ to a realistic scenario in
section~\ref{sec:integrated-born-phase-space}.

A large fraction of the calculation can be simply recycled from the
$e^+e^-$ results presented in section~\ref{sec:recape+e-} with
$L=L_\text{jet}$.
This includes all the contributions involving only collinear physics
(including running-coupling corrections), as a consequence of the
universality of the collinear limit of QCD.
We are therefore left with the computation of the contributions
depending on emissions at large angles (commensurate with the jet
radius).
In general, these corrections depend on the overall structure of the
hard process. In practice, we will consider jets in either $Z$+jet or
inclusive jet samples.
However, we keep the discussion below as
generic as possible so as to facilitate possible future extensions to
other processes.

The $e^+e^-$ results in section~\ref{sec:recape+e-} have been presented
separately for quark and gluon hemispheres. In a $pp$ context, jets can also be
considered as quark- or gluon-initiated (as given by the Born-level
hard process) as long
as we are only interested in the contributions involving either soft
gluon emissions and/or collinear branchings.
More concretely, these include the collinear contributions recycled
from our $e^+e^-$ results, as well as new large-angle contributions
which only depend on soft-gluon emissions.
For these contributions we can consider that we work
with a jet of a given flavour, and ultimately get the total
multiplicity by weighting the quark- and gluon-initiated
jet multiplicities by the corresponding Born-level fractions.
Up to our target \NNDL accuracy, the only contribution where the quark/gluon
separation is not obvious is the case of hard matrix-element
corrections, sensitive to emissions which are both hard and
large-angle. We address this case in section~\ref{sec:nndl-hard-me}.

As with the $e^+e^-$ recap of section~\ref{sec:recape+e-}, we start
with the \DL expressions and then proceed with the calculation of the
\NDL and \NNDL corrections.

\subsection{\DL accuracy}\label{eq:dl-pp}

Since the \DL Lund multiplicity only involves soft-and-collinear emissions, the
results for $h_1^{(q,g)}$ for quark and gluon jets can be taken directly from
the corresponding results in $e^+e^-$ events, i.e.\
Eq.~(\ref{eq:final-master-dl}).
The \DL average Lund multiplicity for a given hard
process is thus given by
\begin{equation}
  h_1(\xi) = f_q h_1^{(q)}(\xi) + f_g h_1^{(g)}(\xi),
\end{equation}
with 
\begin{equation}
  \label{eq:h1-pp-ingredients}
  f_i = \frac{\sigma_i}{\sigma_q+\sigma_g},
  \qquad
  \xi = \as L^2
  \quad \text{ and }\quad
  h_1^{(q,g)}(\xi)=h_{1,e^+e^-}^{(q,g)}(\xi),
\end{equation} 
where the quark/gluon fractions $f_{q,g}$ are computed from the
Born-level cross-sections $\sigma_{q,g}$.

\begin{figure}[t]
  \centering
  \includegraphics[scale=1]{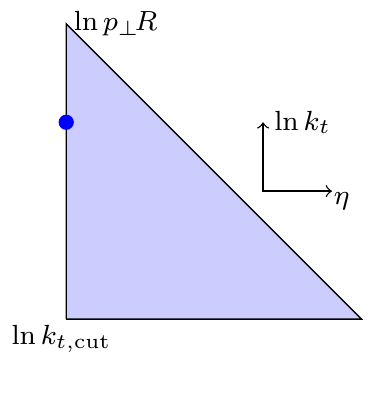}
  \caption{Lund representation of the only new \NDL contribution for high-energy jets: a large-angle emission.}
  \label{fig:ndl-diagram-LA}
\end{figure}

\subsection{\NDL accuracy}
\label{sec:ndl-pp}
\NDL corrections arise from a series of branchings where a single
emission gives a single-logarithmic contribution (proportional to
$\as L$) and all other emissions are soft-and-collinear
(proportional to $\as L^2$).
The two contributions which appear in the $e^+e^-$ case, namely the
hard-collinear and running coupling correction, cf.\
figure~\ref{fig:ndl-ee}, are unchanged in the case of the Lund
multiplicity in a jet.
The only potential new source of \NDL corrections is the case of an
emission at large angles, close to the jet boundary as depicted in
figure~\ref{fig:ndl-diagram-LA}.
Including the proper relative weight of quark- and gluon-initiated
jets, we therefore write
\begin{align}
  \label{eq:h2-decomposition}
  h_2(\xi) = f_q  h_{2, \text{coll}}^{(q)}(\xi) 
  + f_g  h_{2, \text{coll}}^{(g)}(\xi) + h_{2, \text{la}}(\xi).
\end{align}
While $h_{2,\text{la}}$ vanishes in $e^+e^-$
collisions, it is no longer the case
for high-energy jets and we compute this contribution below.

Generically, the radiation of a large-angle gluon depends on
the overall structure of the Born event.
Since the gluon is soft, it is however still possible 
to separate the contribution from quark- and gluon- initiated
jets based on the underlying Born hard process.
We thus write
$h_{2,\text{la}}= f_q h_{2,\text{la}}^{(q)}+f_g h_{2,\text{la}}^{(g)}$.
These contributions would nevertheless be hard-process-dependent, i.e.\ a
``gluon'' jet in a $Z$+jet event would be different from a ``gluon''
jet in an inclusive jet sample.
We proceed by writing the Born hard process as a sum over different
partonic flavour channels:
\begin{equation}
  \sigma^{(i)}
  = \sum_{\substack{\mathcal{C}\in\text{flavour}\\\text{channels}}}
  \sigma_{\mathcal{C}}^{(i)},
\end{equation}
where the (optional) flavour superscript, $i$, limits the sum to
channels contributing to a specific jet flavour.

We start by considering the $\mathcal{O}(\as)$ contribution to
the Lund multiplicity where a single soft gluon is emitted.
For a given channel, the matrix element describing such an emission
can be further decomposed into a sum over coloured dipoles $(ab)$
with $a$ and $b$ two (incoming or outgoing) partons in the Born
process:
\begin{equation}
\label{eq:eikonal}
  \sum_{a,b\in\text{legs}}\frac{\alpha_s}{2\pi} \omega_{ab}^{\mathcal{C}}
  \int k_\perp \dd k_\perp\int \dd y\int\frac{\dd\phi}{2\pi} (k|ab),
  \quad\text{ with }\quad
  (k|ab) = \frac{(p_a\cdot p_b)}{(p_a \cdot k)(k \cdot p_b)} \,,  
\end{equation}
and $\omega_{ab}^{\mathcal{C}}=(-2 \mathbf{T}_a\cdot \mathbf{T}_b)$ is
a colour factor or kinematic weight which, in general, depends on the kinematics of the
full hard event. 
Therefore, the \NDL correction to the Lund multiplicity associated with the
radiation of a soft large-angle gluon takes the form
\begin{equation}
  \label{eq:ndl-la-general}
  \delta N^{(\NDL,i)}_{\text{la}}
  = \sqrt{\as}h_{2, \text{la}}^{(i)}(\xi)
  = \sum_{\substack{\mathcal{C}\in\text{flavour}\\\text{channels}}}
  \frac{\sigma_{\mathcal{C}}^{(i)}}{\sigma^{(i)}}
  \sum_{(ab)\in \text{event}}  \delta N^{(\NDL)}_{\mathcal{C},(ab)},
\end{equation}
where $\delta N^{(\NDL)}_{\mathcal{C},(ab)}$ is the correction to the
multiplicity due to a soft gluon emitted close to the jet boundary
from a dipole $(ab)$. At the first non-trivial order in $\alpha_s$, it
is given by
\begin{equation}
\label{eq:ndl-la-abdip}
  \delta N^{(\NDL)}_{\mathcal{C},(ab),\order{\as}}
  = \frac{\alpha_s}{\pi} \omega_{ab}^\mathcal{C}
    \int_0^{p_\perp}\!\frac{\dd k_\perp}{k_\perp}
    \dd y
    \frac{\dd\phi}{2\pi}
    \left[\frac{k^2_\perp}{2}(k|ab)-\frac{1}{\Delta_k^2}
    \delta_{a\text{ or }b=\text{jet}}\right]
   \Theta(k_t>\ktcut)\, \Theta(\Delta_k<R).
\end{equation}
In this expression, the second term in the square bracket subtracts
the collinear (\DL) contribution and the ``$\delta$'' selects the
cases where one of the two dipole legs corresponds to the jet 
under consideration. This is done so as to isolate the \NDL correction. Also,
$\Delta_k^2= (y - y_\text{jet})^2 + (\phi - \phi_\text{jet})^2$ is the
(squared) distance between the jet axis and the emission.
To evaluate~(\ref{eq:ndl-la-abdip}), we first make a change of
variable from $k_\perp$ to $k_t=k_\perp\Delta_k$. Including the
constraint that $k_t$ has to be above $\ktcut$ we can then write
\begin{equation}\label{eq:from-kperp-to-kt}
  \int_{\ktcut}^{p_\perp\Delta_k} \frac{\dd k_t}{k_t}
  = L +\mathcal{O}(\NNDL),
\end{equation}
where only the logarithmically-enhanced term contributes at \NDL.
This is allowed because the angular integration does not contain any
collinearly-enhanced contribution so that $\Delta_k$ is of order $R$
up to a constant factor which would not generate a logarithmic
enhancement.
We are then left with the integration over the geometric variables $y$
and $\phi$ so that the \NDL correction takes the form
\begin{equation}
\delta N^{(\NDL)}_{\mathcal{C},(ab),\order{\as}}= \frac{\alpha_s}{\pi}L \;\omega_{ab}^\mathcal{C} D_{ab}^{\text{la}}.
\end{equation}
We now compute the coefficients $D_{ab}^{\text{la}}$ for a generic
$2\to 2$ process.
Let us denote the four-momenta of the incoming partons as $p_1$ and
$p_2$, and those of the outgoing jet (for which one measures the
multiplicity) and recoiling parton as $p_j$ and $p_r$.%
\footnote{In the case of $Z$+jet events the recoiling object is the
  colour singlet $Z$ so that $p_r$ does not contribute to the coloured
  dipoles.}
Without loss of generality, we can parametrise the momenta of the
hard legs and the radiated gluon, $k$, as follows (using a
$(p_x,p_y,p_z,E)$ convention):
\begin{subequations}
\begin{align}
\label{eq:param}
  p_1 & \equiv p_\perp \cosh\yhat\,(0,0,1,1),\\
  p_2 & \equiv p_\perp \cosh\yhat\,(0,0,-1,1),\\
  p_j & \equiv p_\perp \,(1,0,\sinh \yhat,\cosh \yhat),\\
  p_r & \equiv p_\perp \,(-1,0,-\sinh \yhat,\cosh \yhat), \\
  k   & \equiv k_\perp \, (\cos \phi, \sin \phi, \sinh y,\cosh y),
\end{align}
\end{subequations}
where $p_\perp$ is the transverse momentum of the jet, $\yhat$ is the
rapidity of the jet in the partonic centre-of-mass frame,\footnote{In
  the lab frame, $\yjet=\yhat+\frac{1}{2}\ln \frac{x_1}{x_2}$
  with $x_{1,2}$ the longitudinal momentum fractions of the incoming partons.}
$k_\perp \ll p_\perp$, and we have picked the jet azimuthal angle to
be zero.
The soft gluon can be emitted from any of the 6 possible colour dipoles:
($p_1,p_2$), ($p_1,p_j$), ($p_2,p_j$), ($p_1,p_r$), ($p_2,p_r$) and ($p_j,p_r$).
Except for the $(12)$ dipole, the integrals in Eq.~\eqref{eq:ndl-la-abdip}
cannot be computed exactly analytically. We thus make a series
expansion in the jet radius $R$ and find~\cite{Lifson:2020gua}
\begin{subequations}\label{eq:large-angle-Rexpansion}
\begin{align}
  D_{12}^{\text{la}} & = \frac{R^2}{2},\\
  D_{1j}^{\text{la}} & = \frac{R^2}{8}+\frac{R^4}{576}+\mathcal{O}(R^8),\\
  D_{jr}^{\text{la}} & = \frac{\tanh^2\yhat}{8}R^2
  +\frac{(\cosh^2\yhat-3)^2}{576\cosh^4\yhat}R^4
  +\frac{\tanh^2\yhat}{384}R^6
  +\mathcal{O}(R^8),\\
  D_{1r}^{\text{la}} & = \frac{e^{2\yhat}}{8\cosh^2\yhat}R^2
  +\frac{1}{64\cosh^4 \yhat}R^4
  +\frac{\tanh^2\yhat}{384\cosh^4\yhat}R^6
  +\mathcal{O}(R^8).
\end{align}
\end{subequations}
Then, $D_{2j}^{\text{la}} = D_{1j}^{\text{la}}$ and $D_{2r}^{\text{la}}$ is
obtained from $D_{1r}^{\text{la}}$ taking $\yhat \to -\yhat$.  One
notices that the $(1r)$ dipole tends to the $(12)$ dipole when
$\yhat\to\infty$ and to 0 when $\yhat\to -\infty$.  Similarly, the
$(jr)$ dipole is symmetric under $\yhat \to -\yhat$ and converges to
the $(1j)$ dipole at large $\yhat$.
Dipoles involving the recoiling parton depend on the kinematics of the
$2\to 2$ process via $\yhat$. This will have to be taken into account
when considering jets within a rapidity range (see
section~\ref{sec:integrated-born-phase-space}).
In practice, we have checked numerically that the small-$R$ expansion
used in Eq.~\eqref{eq:large-angle-Rexpansion} reproduces the exact
result to well within 1\% up to $R=1$, and up to deviations of at most
0.3\% for $R=0.8$ which we use as a default below.

At all orders, one should include the extra contribution to the
multiplicity due to subsequent soft-and-collinear radiation from the large-angle
gluon. Using Eq.~\eqref{eq:n-ndl-full}, we find that ($\nu = \sqrt{\abar L^2}$)
\begin{equation}\label{eq:NDL-large-angle-result}
  \delta N^{(\NDL)}_{\mathcal{C},(ab)}
  = \frac{\alpha_s}{\pi} \omega_{ab}^\mathcal{C} D_{ab}^{\text{la}} \int_0^L d\ell\, N_g^{(\DL)}(L-\ell)
  = \sqrt{\frac{\alpha_s}{2C_A\pi} }\; \omega_{ab}^\mathcal{C} D_{ab}^{\text{la}} \,\sinh\nu.
\end{equation}
To compute the \NDL multiplicity for a given hard process, one should
also specify its corresponding weights $\omega_{ab}^\mathcal{C}$.
In the case of $Z$+jet events there are only two partonic channels,
one corresponding to quark jets and the other to gluon jets. Since
there are only three coloured partons (two incoming and one outgoing),
the colour algebra for each channel is straightforward and the
$\omega_{ab}^\mathcal{C}$ weights are listed in
table~\ref{tab:Born-Zj}.
The situation is more involved for dijet events due to both the
increased number of partonic channels and the more intricate colour
structure of the four-leg process.
The corresponding $\omega_{ab}^\mathcal{C}$ weights can be
extracted from Ref.~\cite{Ellis:1986bv} and are given in
appendix~\ref{app:dijets} for completeness.

\begin{table}
  \begin{center}
    \begin{tabular}{l c c c }
      \toprule
      Channel
      & $\omega_{12}$
      & $\omega_{1j}$
      & $\omega_{2j}$\\
      \midrule
      $q\bar q\to Zg$
      & $2C_F - C_A$
      & $C_A$
      & $C_A$\\
      $qg \to Zq$
      & $C_A$
      & $C_A$
      & $2C_F -C_A$\\
      \bottomrule
    \end{tabular}
  \end{center}
  \caption{Colour coefficients $\omega_{ab}^{\mathcal{C}}\equiv (-2 \mathbf{T}_a\cdot \mathbf{T}_b)$ 
   for different flavour channels in $Z+$jet events.}\label{tab:Born-Zj} 
\end{table}

\subsection{\NNDL accuracy}\label{sec:NNDL-accuracy}

\begin{figure}
   \begin{subfigure}[t]{0.32\linewidth}
   \centering
    \includegraphics[width=0.75\textwidth]{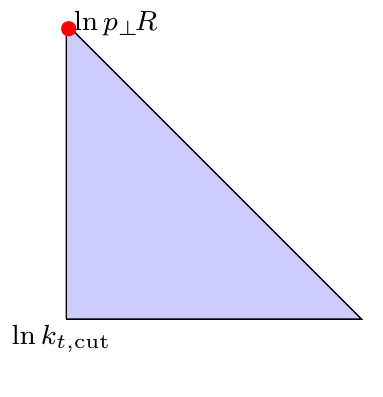}
    \caption{Hard matrix-element.}
    \label{fig:nndl-diagram-ME}
  \end{subfigure}
   \begin{subfigure}[t]{0.32\linewidth}
   \centering
    \includegraphics[width=0.75\textwidth]{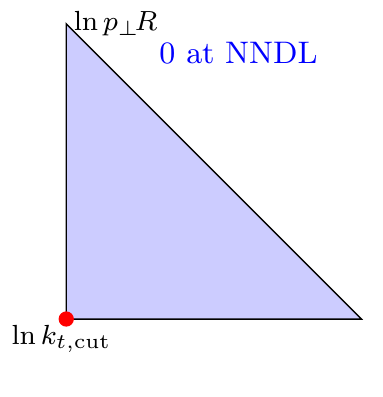}
    \caption{Large-angle, $k_t\sim \ktcut$.}
    \label{fig:nndl-diagram-LA-ktcut}
  \end{subfigure}
   \begin{subfigure}[t]{0.32\linewidth}
   \centering
    \includegraphics[width=0.75\textwidth]{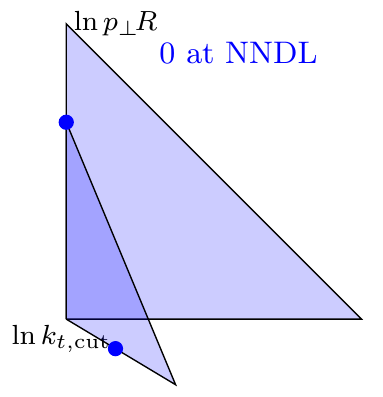}
    \caption{Large-angle $\times$ $k_t\sim \ktcut$.}
    \label{fig:nndl-diagram-LAxL}
  \end{subfigure}
  \\[3mm]
  \begin{subfigure}[t]{0.32\linewidth}
  \centering
    \includegraphics[width=0.75\textwidth]{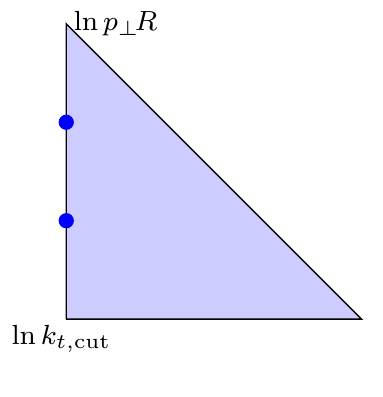}
    \caption{Two large-angle.}
    \label{fig:nndl-diagram-LA}
  \end{subfigure}
  \begin{subfigure}[t]{0.32\linewidth}
  \centering
    \includegraphics[width=0.75\textwidth]{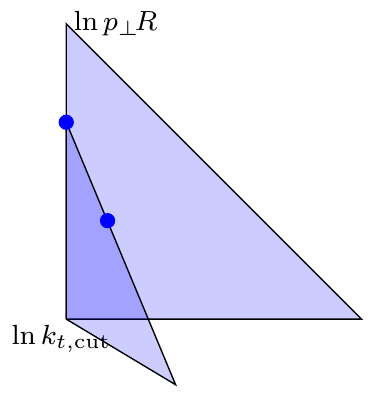}
    \caption{Large-angle $\times$ hard-collinear.}
    \label{fig:nndl-diagram-LAxhc}
  \end{subfigure}
  \begin{subfigure}[t]{0.32\linewidth}
  \centering
    \includegraphics[width=0.75\textwidth]{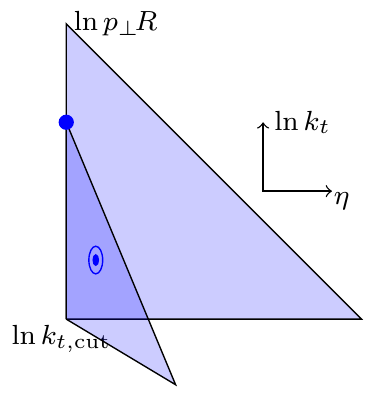}
    \caption{Large-angle $\times$ running $\as$.}
    \label{fig:nndl-diagram-LAxrc}
  \end{subfigure}
  \caption{Lund representation of the new configurations 
    with respect to $e^+e^-$ collisions that 
    contribute to the multiplicity at \NNDL.
  }\label{fig:nndl-diagrams}
\end{figure}

At \NNDL accuracy, we again have to include two broad families of contributions:
collinear contributions which can be obtained from the corresponding
calculation in $e^+e^-$, namely $h_{3,\text{coll}}$ given by
Eq.~\eqref{eq:h3-ee}, as well as extra corrections involving
large-angle emissions.
The latter are depicted in figure~\ref{fig:nndl-diagrams} and we
briefly discuss them here before proceeding with their explicit calculation in
the following sections.

\NNDL corrections can be obtained in two ways (see the discussion in
section~\ref{sec:recape+e-}): either from one ``\NNDL-like'' emission
contributing a factor $\as$ (with no logarithmic enhancement)
accompanied by any number of soft-and-collinear emissions, or from two
``\NDL-like'' emissions, each contributing a factor $\as L$, plus any
number of soft-and-collinear emissions.
In the first case, contributions at large angles can arise from two
kinematic configurations:
\begin{itemize}[noitemsep,topsep=2pt,parsep=2pt]
\item an emission at the top of the Lund plane,
  figure~\ref{fig:nndl-diagram-ME}, corresponding to a hard
  matrix-element correction, discussed in
  section~\ref{sec:nndl-hard-me},
\item or an emission at the bottom-left corner of the Lund plane,
  figure~\ref{fig:nndl-diagram-LA-ktcut}, corresponding to an emission 
  at large angle and close to the $\ktcut$ kinematic limit.
  We have structured the calculation so that this contribution is
  zero, as discussed in section~\ref{sec:nndl-top-v-bottom}.
\end{itemize}
In the second case, the first \NDL-like emission must happen at large
angle, and the second one can correspond to any \NDL correction. The
second emission can therefore be
\begin{itemize}[noitemsep,topsep=2pt,parsep=2pt]
\item a collinear emission close the the kinematic $\ktcut$,
  figure~\ref{fig:nndl-diagram-LAxL}, which also yields a vanishing
  contribution as discussed in section~\ref{sec:nndl-top-v-bottom},
\item another large-angle emission, figure~\ref{fig:nndl-diagram-LA},
  discussed in section~\ref{sec:nndl-two-la},
\item a hard-collinear branching, figure~\ref{fig:nndl-diagram-LAxhc},
  discussed in section~\ref{sec:nndl-la-hc},
\item or a running-coupling correction,
  figure~\ref{fig:nndl-diagram-LAxrc}, discussed in
  section~\ref{sec:nndl-la-rc}.
\end{itemize}
Altogether, the \NNDL function $h_3(\xi)$ for the Lund multiplicity in
high-energy jets is written as\footnote{In this expression, the quark
  and gluon fractions are taken from the Born-level process as in 
  the \DL and \NDL sections. At \NNDL accuracy, NLO corrections to the Born-level
  cross-sections have to be taken into account and we include them in
  $h_{3,\text{hme}}(\xi)$ as detailed in section~\ref{sec:nndl-hard-me}.}
\begin{align}
  \label{eq:nndl-function-definition}
  h_3(\xi)
  = f_q h_{3, \text{coll}}^{(g)}(\xi)
  + f_g h_{3, \text{coll}}^{(q)}(\xi)
  + h_{3,\text{hme}}(\xi)
  + h_{3,\text{la}^2}(\xi)
  + h_{3,\text{la}\times\text{hc}}(\xi)
  + h_{3,\text{la}\times\text{rc}}(\xi),
\end{align}
where the collinear $h_{3, \text{coll}}^{(i)}(\xi)$ were defined in
Eqs.~\eqref{eq:h3-coll-q}, \eqref{eq:h3-coll-g}, and we discuss
the calculation of the final four pieces in the following sections.
%

\subsubsection{Hard matrix-element correction (top of the Lund plane)}
\label{sec:nndl-hard-me}

We first consider the \NNDL correction to the Lund multiplicity of a
hard, wide-angle emission that arises due to the fact that the soft
eikonal approximation is no longer valid and the exact matrix element
has to be used instead.
Our starting point is the exact expression for the Lund multiplicity
at NLO, i.e.\ $\order{\as}$ relative to the Born-level process
\begin{equation}
\label{eq:nndl-hme-as}
\avnlp_{\order{\as}} = \frac{1}{\sigma_0+\sigma_1} \left(\sigma_0+ \int \dd \Phi  |\mathcal M_R|^2 [1 + \Theta(k_{t}>\ktcut) ]
+ \int \dd \Phi  |\mathcal M_V|^2 \right),
\end{equation}
where $\sigma_0$ and $\sigma_1$ are respectively the Born-level cross section and
its NLO correction, and $|\mathcal M_{R,(V)}|$ is the real (virtual) NLO 
matrix-element for the process being studied.
For the real contribution, we have assumed that the two partons are
inside the jet under consideration. The generalisation to the case where
only one parton is inside the jet is straightforward and discussed below.

The extension of Eq.~\eqref{eq:nndl-hme-as} to all orders requires
further dressing of each term with towers of \DL emissions.
Since this depends on the partonic flavour, one has to separate each
term in~(\ref{eq:nndl-hme-as}) into flavour channels.  
Let us therefore assume that one can associate a suitably-defined
flavour index $a$ with the virtual contribution, and two flavour indices
$b$ and $c$ with the two partons in the real contribution, corresponding
to an $a\to bc$ splitting.
For the sake of the argument, let us further assume that
we can separate the real and virtual contributions into a ``hard'' and
a ``soft'' piece:
\begin{align}
  |\mathcal M_R^{(bc)}|^2 & = |\mathcal M_R^{(bc)}|^2_\text{hard} + |\mathcal M_R^{(bc)}|^2_\text{soft},\\
  |\mathcal M_V^{(a)}|^2 & = |\mathcal M_V^{(a)}|^2_\text{hard} + |\mathcal M_V^{(a)}|^2_\text{soft}.
\end{align}
For the real contribution, one can define
$|\mathcal M_R^{(bc)}|^2_\text{soft}$ so that it contains the soft
and/or collinear divergences which reproduce the contributions to the
multiplicity not associated with hard-wide-angle physics.  This
includes the \DL and \NDL contributions, the \NNDL collinear-endpoint
correction (which can be recycled from the $e^+e^-$ result and
corresponds to the terms involving the $D^{a\to bc}_\text{end}$ coefficients in
Eqs.~(\ref{eq:h3-coll-q}) and~(\ref{eq:h3-coll-g})) and potential
\NNDL soft-large-angle corrections (discussed in
section~\ref{sec:nndl-top-v-bottom}).  This leaves only a non-zero
$|\mathcal M_R^{(bc)}|^2_\text{hard}$ contribution at the top of the
Lund plane.
Such a separation is less obvious for the virtual correction, but, due
to infrared safety, one can take
$|\mathcal M_V^{(a)}|^2_\text{soft}=-|\mathcal
M_R^{(bc)}|^2_\text{soft}$ for an $a\to bc$ splitting.
In practice, the \NNDL hard matrix-element contribution to the
multiplicity only requires the integrated cross-sections
$\sigma_{R(V),\text{hard(soft)}}=\int d\Phi  |\mathcal M_{R(V)}|^2_\text{hard(soft)}$.

The all-orders average multiplicity including all hard matrix-element
effects can then be written as
\begin{align}
\label{eq:nndl-hme-2}
  \avnlp = \frac{1}{\sigma_0+\sigma_1}&\left\{
    \sum_a \sigma_0^{(a)} N_a(L)
    +\sum_{bc} \sigma_{R,\text{hard}}^{(bc)} [N_b(L)+N_c(L)]\right. \nonumber \\
    &\left.+\sum_{bc} \sigma_{R,\text{soft}}^{(bc)}\delta_{a=b+c} N_a(L)
    +\sum_{a} \sigma_{V}^{(a)} N_a(L) \right\}.
\end{align}
In writing this expression, we have taken into account the fact that
for the \DL dressing of the real ``soft'' contribution, one has to
define the flavour sum $a=b+c$ as a proxy for the flavour of the
parent parton corresponding to the $(bc)$ flavour pair. The exact
procedure for the reconstruction the flavour of $a$ from the daughter
partons is irrelevant as long as it is correct in the soft limit.

Adding and subtracting $\sigma_{R,\text{hard}}^{(bc)}\delta_{a=b+c} N_a(L)$
to Eq.~\eqref{eq:nndl-hme-2} we obtain
\begin{align}
\label{eq:nndl-hme-3}
\avnlp = \frac{1}{\sigma_0+\sigma_1}&\left\{
    \sum_a \sigma_0^{(a)} N_a(L)
    +\sum_{bc} \sigma_{R,\text{hard}}^{(bc)} [N_b(L)+N_c(L) - N_a(L)]\right.\nonumber \\
    & \left.+\sum_{a} \left[\sum_{bc} \sigma_{R}^{(bc)}\delta_{a=b+c} + \sigma_{V}^{(a)}\right]N_a(L) \right\},
\end{align}
where each term can be determined by means of fixed-order Monte Carlo
simulations which provide access to the flavours of the partons.
The first term is obtained from the quark and gluon contributions to
the Born-level cross section. As we have already mentioned, the second
term is calculated from the real NLO correction to the
cross-section. To isolate the ``hard'' component, we run with a small
$\ktcut$ and subtract the $\mathcal{O}(\as)$ contributions not
associated with hard-wide-angle physics, i.e.\ the \DL and \NDL
contributions as well as the collinear endpoint \NNDL contribution,
figure~\ref{fig:nndl-diagram-hccut}.
The \DL-dressing factor $N_b(L)+N_c(L)-N_a(L)$ can be interpreted as
if a hard parton of flavour $a$ --- which would have been dressed with a
\DL multiplicity $N_a(L)$ --- is replaced by two hard partons $b$ and
$c$, dressed with \DL multiplicities $N_b(L)$ and $N_c(L)$
respectively.
Finally, the third term in Eq.~\eqref{eq:nndl-hme-3} can simply be
interpreted as the change in quark and gluon fractions at NLO,
including both real and virtual corrections.
While the individual contributions from the second and third terms depend on
the precise recipe to define the combined flavour of a flavour
pair $(bc)$, their sum is well-defined.

Let us briefly return to the question of what happens if in a
$2\to 3$ real event each of the three partons are in separate jets. In
this case, the second term in Eq.~\eqref{eq:nndl-hme-3} disappears
(imagine $b$ is in the jet, we have $a=b$ and the square bracket
becomes $N_b-N_a=0$). The last term would then just have one of $b$ or
$c$ in the jet which corresponds to the NLO correction to the jet
cross-section.

We conclude this discussion by summarising the recipe for obtaining the
\NNDL hard matrix-element corrections to the Lund multiplicity.
\begin{enumerate}[noitemsep,topsep=2pt,parsep=2pt]
\item Our prescription to define the jet flavour is as follows. If the
  jet has a single parton, the jet flavour is taken as the flavour of
  that parton. If instead the jet has 2 partons of flavour indices
  $b,c$, the sum $b+c$ is defined as follows: for 2 gluons the sum is
  a gluon, for a quark and a gluon the sum is a quark, for a
  same-flavour $q\bar q$ pair the sum is a gluon, and for all the
  other cases the sum is taken as a quark.
\item The Born-level cross-sections $\sigma_0^{(a)}$ for quark and
  gluon flavours can be obtained from a Monte Carlo simulation
  including all the required fiducial cuts.
\item With the above flavour prescription, the NLO corrections
  $\sigma_1^{(a)}$ for quark and gluon jets can be obtained, including
  fiducial cuts, from any Monte Carlo generator providing information
  about the flavour of the final-state partons.
  This defines
  \begin{equation}
    \frac{\alpha_s}{2\pi}D_\text{hme,NLO}^{(a)} = \frac{\sigma_1^{(a)}}{\sigma_0^{(a)}}.
  \end{equation}
\item We compute the second term in~(\ref{eq:nndl-hme-3}) from a
  simulation of the 
  real NLO corrections to the cross-section. We select jets with 
  two partons (of flavour $b$ and $c$) and study the cumulative 
  distribution of their relative $\ln k_t$.
  For each flavour channel, after subtracting the known \DL and \NDL
  contributions, the cumulative distribution tends to a constant at
  small $\ln k_t$. Further subtracting the
  \NNDL contribution from the collinear endpoint from this constant
  gives $\sigma_{R,\text{hard}}^{(bc)}$.
  We then define
  \begin{equation}
    \frac{\alpha_s}{2\pi} D_{\text{hme},R}^{(bc)} = 
    \frac{\sigma_{R,\text{hard}}^{(bc)}}{\sigma_0^{(b+c)}}.
  \end{equation}
  A practical example of the extraction of $D_{\text{hme},R}^{(bc)}$
  from Monte Carlo simulations is given in
  appendix~\ref{app:D-hme-bc-extraction}.
\end{enumerate}
The last hard matrix-element correction that has to be taken into account arises
by expanding the normalisation in~(\ref{eq:nndl-hme-3}) as 
$\frac{1}{\sigma_0+\sigma_1} \sim \frac{1}{\sigma_0}\Big(1 - \frac{\sigma_1}{\sigma_0}\Big)$.
The \NNDL hard matrix-element correction to the Lund multiplicity
thus takes the form
\begin{align}
  \delta N^{(\NNDL)}_{\text {hme}}
  & =
  \sum_a \left(\frac{\alpha_s}{2\pi}D_\text{hme,NLO}^{(a)}-\frac{\sigma_1}{\sigma_0}\right)
  f_a N^{(\DL)}_a(L) \nonumber\\
  & + \sum_{bc} f_{b+c}\frac{\alpha_s}{2\pi}D_{\text{hme},R}^{(bc)}[N^{(\DL)}_b(L)+N^{(\DL)}_c(L)-N^{(\DL)}_a(L)],
\end{align}
with $f_a=\frac{\sigma_0^{(a)}}{\sigma_0}$ the Born-level partonic
fractions.
One can gather the quark and gluon contributions and write
\begin{equation}
  \as h_{3,\text{hme}}(\xi) = \delta N^{(\NNDL)}_{\text {hme}}= \frac{\alpha_s}{2\pi}\left[D_\text{hme}^{q}N^{(\DL)}_q(L) + D_\text{hme}^{g}N^{(\DL)}_g(L)\right],
\end{equation}
with
\begin{subequations}
\begin{align}
  D_\text{hme}^{q}
  &= \left(D_\text{hme,NLO}^{(q)}+D_{\text{hme},R}^{(qg)}+D_{\text{hme},R}^{(qq)}-\frac{2\pi\sigma_1}{\alpha_s\sigma_0}\right)f_q
     + 2D_{\text{hme},R}^{(q\bar q)}f_g,\\
  D_\text{hme}^{g}
  &=\left(D_\text{hme,NLO}^{(g)}+D_{\text{hme},R}^{(gg)}-D_{\text{hme},R}^{(q\bar q)}-\frac{2\pi\sigma_1}{\alpha_s\sigma_0}\right)f_g.
\end{align}
\end{subequations}

\subsubsection{Interplay between the top and bottom of the Lund
  plane}\label{sec:nndl-top-v-bottom}

Let us now focus on the contribution associated with
diagram~\ref{fig:nndl-diagram-LA-ktcut}, corresponding to an endpoint
contribution for the radiation of a soft gluon at large angles.
Going back to the derivation of the \NDL large-angle contribution from
section~\ref{sec:ndl-pp}, we note that, after we changed variable
from $k_\perp$ to $k_t=k_\perp\Delta$, the integration over $k_t$
gave both the \NDL contribution and an \NNDL correction. More
precisely, Eq.~(\ref{eq:from-kperp-to-kt}) can be rewritten as
\begin{equation}\label{eq:ndl-la-nndl-top}
  \int_{\ktcut}^{p_\perp\Delta_k} \frac{\dd k_t}{k_t}
   = \int_{\ktcut}^{p_\perp R} \frac{\dd k_t}{k_t}
   - \int_{p_\perp\Delta_k}^{p_\perp R} \frac{\dd k_t}{k_t}
  = L + (\text{NNDL}_\text{top}),
\end{equation}
where we see that the \NNDL correction is associated with the hard $k_t$
scale and can therefore be absorbed into the hard matrix-element
correction computed in the previous section.
As a corollary, the contribution at the bottom of the Lund plane,
diagram~\ref{fig:nndl-diagram-LA-ktcut}, vanishes.

The argument can be extended to the case of
diagram~\ref{fig:nndl-diagram-LAxL}, where a first large-angle
correction would be followed by a secondary emission close to the
$\ktcut$ threshold.
For this, we consider the all-order expression obtained by adding a
tower of soft-collinear gluons to a first large-angle emission as
given by Eq.~\eqref{eq:ndl-la-abdip} (assuming for simplicity that the
$(ab)$ dipole does not involve the jet leading parton):
\begin{equation}
  \delta N_{\mathcal{C},(ab),\order{\as}}
  = \frac{\alpha_s}{2\pi} \omega_{ab}^\mathcal{C}
    \int_0^{p_\perp}\!\frac{\dd k_\perp}{k_\perp}
    \int_{\Delta_k<R} \! \dd y\,
    \frac{\dd\phi}{2\pi}\,
    k^2_\perp(k|ab)
   \Theta(k_t>\ktcut)N^{(\DL)}\!\Big(L;\ln\frac{p_\perp R}{k_\perp\Delta_k}\Big),
\end{equation}
where the last factor includes soft-collinear gluon radiation from the
$k_t$ scale of the first emission, $k_\perp\Delta_k$, down to the cut-off
scale $\ktcut$.
Changing variables from $k_\perp$ to $k_t=k_\perp\Delta_k$, one recovers
the same picture as above, namely the \NDL contribution plus an \NNDL
correction when the first emission is at the top of the Lund plane
(the second term in Eq.~\eqref{eq:ndl-la-nndl-top}) dressed with \DL
soft-collinear radiation.
In particular, written in terms of $k_t$, the dressing factor
$N^{(\DL)}\!\left(L;\ln\frac{p_\perp R}{k_\perp\Delta_k}\right)$ is
simply $N^{(\DL)}\!\left(L;\ln\frac{p_\perp R}{k_t}\right)$, i.e.\ the
\DL result with no \NDL corrections.
This means that the contribution from
diagram~\ref{fig:nndl-diagram-LAxL} also vanishes.

This vanishing of
diagrams~\ref{fig:nndl-diagram-LA-ktcut}
and~\ref{fig:nndl-diagram-LAxL} carries some degree of arbitrariness
and indicates a freedom in the calculation 
to reshuffle contributions across different corrections.
More concretely, a fraction of the contribution  
associated with the top of the Lund plane can be
redistributed into a contribution at the bottom of the Lund plane
(starting at $\mathcal{O}(\as)$) and a contribution associated with
diagram~\ref{fig:nndl-diagram-LAxL} (starting at
$\mathcal{O}(\as^2)$). This reorganisation is done without changing
the all-order \NNDL final result.\footnote{A similar effect is present
  in the collinear limit where there is some freedom in reshuffling
  contributions between the top of the Lund plane, the collinear
  endpoint and a hard-collinear emission with a second emission at
  commensurate angles, see section~4.3.3 of
  Ref.~\cite{Medves:2022ccw}.}
To illustrate this, let us take a different approach where we instead  
change variable from $k_\perp$ to, e.g., $\tilde k_\perp = k_\perp R$.
In this case, the integration with a single large-angle gluon, i.e.\
the analogue of Eq.~\eqref{eq:ndl-la-nndl-top}, would
give
\begin{equation}\label{eq:from-kperp-to-ktilde}
  \int_0^{p_\perp R} \frac{\dd \tilde k_\perp}{\tilde k_\perp}\Theta(k_t>\ktcut) 
   =   
  \int_{\ktcut\frac{\Delta_k}{R}}^{p_\perp R} \frac{\dd k_t}{k_t}
  = \int_{\ktcut}^{p_\perp R} \frac{\dd k_t}{k_t}
  + \int_{\ktcut \Delta_k/R}^{\ktcut} \frac{\dd k_t}{k_t}
  = L +(\text{NNDL}_{\tilde k_\perp,\text{bottom}}),
\end{equation}
with
\begin{equation}
(\text{NNDL}_{\tilde k_\perp,\text{bottom}}) =
(\text{NNDL}_\text{top})
= \ln\frac{\Delta_k}{R},
\end{equation}
with $(\text{NNDL}_\text{top})$ the contribution from the top of the
Lund plane from Eq.~\eqref{eq:ndl-la-nndl-top}.
After integrating over $\Delta_k$, this would give a constant.
At order $\as$, this is simply equivalent to reshuffling
an \NNDL correction from the top to the bottom of the Lund plane.
At all orders, the contributions from the top and bottom of the Lund
plane are dressed with different \DL contributions.
In practice, one can show that since the contribution from
diagram~\ref{fig:nndl-diagram-LAxL} is also non-zero, the
all-order result remains the same independently of the use of 
$\tilde k_\perp$ or $k_\perp$, as expected.
%
\subsubsection{Two large-angle emissions}
\label{sec:nndl-two-la}

We now investigate the case in which a Born-level configuration
$(p_1p_2\to p_j p_r)$ emits a soft-and-large-angle gluon, $k_1$,
which then radiates a soft emission $k_2$ at a commensurate angle,
$\Delta_{k_1 k_2} \sim \Delta_{j k_1}$.
Due to the purely soft nature of this correction, we can again split it
in a quark- and a gluon-jet contribution
\begin{equation}
  \label{eq:ndl-la2-general}
  h_{3,\text{la}^2}(\xi) = f_q h_{3, \text{la}^2}^{(q)}(\xi) + f_g h_{3, \text{la}^2}^{(g)}(\xi)
  \quad
  \text{with}
  \quad
  \as h_{3, \text{la}^2}^{(i)}(\xi) 
  = \sum_{\mathcal{C}}
  \frac{\sigma_{\mathcal{C}}^{(i)}}{\sigma^{(i)}}
  \sum_{(ab)\in \text{event}}  \delta N^{(\NNDL)}_{\mathcal{C},(ab),\text{la}^2},
\end{equation}
where we decomposed the hard Born-level process into a sum over partonic
channels, $\mathcal{C}$, and coloured dipoles $(ab)$, as in
section~\ref{sec:ndl-pp}.
While the methods in section~\ref{sec:ndl-pp} allow us to account for
the primary emission at full colour, we here adopt a simplified
approach and describe the second emission in the large-$N_c$
limit. If the gluon $k_1$ is radiated from
dipole $(ab)$, we thus only have to consider, for a real $k_1$, the radiation of gluon $k_2$ from
either the $(ak_1)$ or the $(k_1b)$ dipole and, for a virtual $k_1$,
the radiation of $k_2$ from the $(ab)$ dipole.
At $\order{\as^2}$, the Lund multiplicity is thus given by
\begin{align}
  \label{eq:nndl-la2-0}
  \delta N_{\mathcal C,(ab),\text{la}^2,\order{\as^2}}
  & = \left(\frac{\as}{2\pi}\right)^2 \omega_{ab}^{\mathcal C} N_c
    \int_0^{p_\perp}\!  \frac{\dd k_{\perp 1}}{k_{\perp 1}}
    \int_0^{k_{\perp 1}}\! \frac{\dd k_{\perp 2}}{k_{\perp 2}}
    \int\!\dd y_1  \dd y _2 
     \frac{\dd \phi_1}{2\pi} \frac{\dd \phi_2}{2\pi}  \\
  &  \times k_{\perp 1}(k_1|ab)\:
    k_{\perp 2}[(k_2|a k_1) + (k_2| b k_1) - (k_2 | ab)] \,
    \Theta(\Delta_2 < R)  \Theta(k_{t2} > \ktcut), \nonumber
\end{align}
with $\Delta_i$ the distance between gluon $k_i$ and the jet axis.
Here, the eikonal factors were defined in Eq.~\eqref{eq:eikonal},
$\Theta(\Delta_2 < R)$ ensures that $k_2$ is emitted inside of the
jet, and we made explicit the use of the large-$N_c$ approximation for
the second emission. Note that the gluon $k_1$ can be emitted either
inside or outside of the jet.
In writing Eq.~(\ref{eq:nndl-la2-0}), we have implicitly assumed that
the jet had been clustered with the anti-$k_t$ algorithm, allowing us
to write the condition that the second emission is in the jet as
$\Theta(\Delta_2<R)$. For a different jet algorithm, this would have
to be adapted, hence potentially yielding a different result at
\NNDL accuracy.

\begin{figure}
  \begin{minipage}{0.45\textwidth}
    \includegraphics[width=\textwidth]{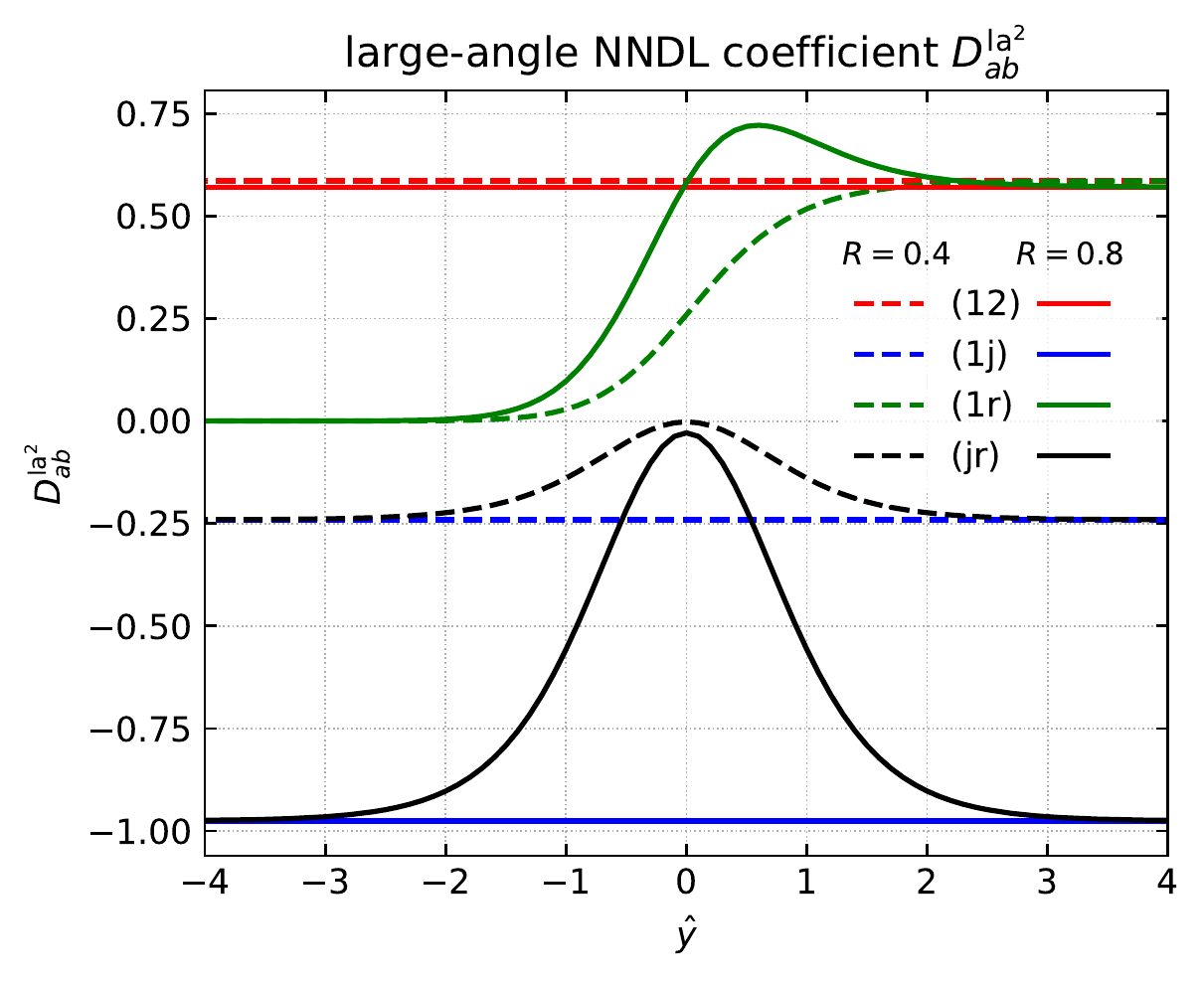}
  \end{minipage}
  \hfill
  \begin{minipage}{0.52\textwidth}
    \begin{center}
      \begin{tabular}{c c c}
        \multicolumn{3}{c}{Values for $\yhat=0$}\\
        \toprule
        $R$ & 0.4 & 0.8\\
        \midrule
        $D_{12}^{\text{la}^2}$ & 0.5864261(12) & 0.5712475(12) \\
        $D_{1j}^{\text{la}^2} = D_{2j}^{\text{la}^2}$
            & -0.24098(15) & -0.97564(17) \\
        $D_{1r}^{\text{la}^2} = D_{2r}^{\text{la}^2}$
            & 0.259460(11) & 0.582097(13) \\
        $D_{jr}^{\text{la}^2}$
            & -0.0017776(17) & -0.0285608(17)\\
        \bottomrule
      \end{tabular}
    \end{center}
  \end{minipage}
  \caption{\NNDL coefficients $D_{ab}^{\text{la}^2}$ for each dipole
    coming from two emissions at large angles. The results are
    provided for both $R=0.4$ and $R=0.8$ jets. The table on the right
    gives numerical values for $\yhat=0$.}\label{fig:la2-coeff}
\end{figure}

In order to isolate the \NNDL terms in Eq.~\eqref{eq:nndl-la2-0}, we
first subtract the \DL and \NDL contributions.
For dipoles not involving the jet ($(12)$, $(1r)$ and $(2r)$),
$(k_1|ab)$ has no collinear singularity when $\Delta_1 \to 0$.
The only contribution up to \NDL is thus that of $k_1$ being emitted
inside the jet at large angles with $k_2$ being a \DL
soft-and-collinear dressing. One therefore has the following
subtracted integrand:
\begin{equation} \label{eq:I-ab-nojet}
  I_{ab} = 
    k_{\perp 1}^2(k_1|ab)\:
    \left\{k_{\perp 2}^2[(k_2|a k_1) + (k_2| k_1b) - (k_2 | ab)]
    \Theta(\Delta_2 < R) 
  - \frac{4}{\Delta^2_{12}}\Theta(\Delta_1, \Delta_{12} < R)\right\}.
\end{equation}
For dipoles involving the jet ($(1j)$, $(2j)$ and $(rj)$), there are
two contributions to subtract: a \DL one where both gluons are soft
and collinear with the second radiated as a secondary emission from
the first one, and a \NDL one where $k_1$ occurs at large angles and
$k_2$ is a \DL soft-and-collinear dressing. This gives
\begin{align}\label{eq:I-ab-jet}
  I_{aj}
  &=  
    k_{\perp 1}^2 (k_1|1j)
    k_{\perp 2}^2[(k_2|a k_1) + (k_2| j k_1) - (k_2 | aj)]
    \Theta(\Delta_2 < R)  \\
  &- \frac{2}{\Delta^2_1}\frac{4}{\Delta^2_{12}} \Theta(\Delta_{12}<\Delta_1<R)
    - \Big[k_{\perp 1}^2(k_1|aj)-\frac{2}{\Delta^2_1}\Big]\frac{4}{\Delta^2_{12}} 
    \Theta(\Delta_1, \Delta_{12}<R). \nonumber
\end{align}
In writing the subtraction terms, we have split the radiation of the
first soft gluon into a collinear contribution $2/\Delta_1^2$ and a
large-angle contribution $k_{\perp 1}^2(k_1|aj)-2/\Delta^2_1$.
After performing these subtractions, the angular integrations are finite and do
not show any collinear enhancement. At $\mathcal{O}(\as^2)$ we
can thus trivially perform the $k_\perp$ integrals by replacing
$k_{t2} \to k_{\perp2}R$ (see Eq.~\eqref{eq:from-kperp-to-kt}), which
simply gives a factor $L^2/2$. The \NNDL correction to the
multiplicity is then
\begin{align}
  \label{eq:nndl-la2-1}
  \delta N_{\mathcal C,(ab),\text{la}^2,\order{\as^2}}^{(\NNDL)} 
  =
  \left(\frac{\as}{2\pi}\right)^2 \omega^{\mathcal C}_{ab} N_c \frac{L^2}{2} 
  D_{ab}^{\text{la}^2},
\end{align}
with
\begin{align}
  \label{eq:nndl-la2-D}
  D_{ab}^{\text{la}^2}
  =
  \int_{-\infty}^\infty\! \dd y_1 \int_{-\infty}^\infty\! \dd y _2 
  \int_{-\pi}^\pi \frac{\dd \phi_1}{2\pi}\int_{-\pi}^\pi \frac{\dd \phi_2}{2\pi} 
  \; I_{ab}.
\end{align}
We could not find analytic expressions for the integrals in
Eq.~\eqref{eq:nndl-la2-D} so we evaluated them
numerically.\footnote{We could tentatively do an expansion in the jet
  radius $R$ as in section~\ref{sec:ndl-pp} but the coefficients of
  this expansion would also have to be evaluated numerically so we
  preferred the fully numerical solution.
  We note however that, at small $R$, the $D^{\text{la}^2}_{ab}$
  coefficients behave like $R^2\ln R^2$. This behaviour comes from
  configurations where the first emission is outside the jet but very
  close to the jet boundary with the second emission inside the jet
  and also close to the jet boundary so that the $\ln R^2$ enhancement
  is of collinear origin. See
  e.g.~\cite{Dasgupta:2012hg,Hatta:2017fwr} for additional
  discussions.  }
As for the \NDL $D^{\text{la}}_{ab}$ coefficients, the
$D^{\text{la}^2}_{ab}$ coefficients involving the recoiling jet depend
on the rapidity difference between the jet and the recoiling jet,
$2\yhat$ in the notations of section~\ref{sec:ndl-pp}.
For reproducibility purposes, we show the coefficients $D^{\text{la}^2}_{ab}$ in
figure~\ref{fig:la2-coeff} together with the numerical values obtained
for $\yhat=0$ for two different jet radii.
Furthermore, $D_{2j}^{\text{la}^2} = D_{1j}^{\text{la}^2}$ and $D_{2r}^{\text{la}^2}$ is
obtained from $D_{1r}^{\text{la}^2}$ taking $\yhat \to -\yhat$. As
expected, $D_{1r}^{\text{la}^2}$ goes to 0 at large 
negative $\yhat$ and to $D_{12}^{\text{la}^2}$ at large positive
$\yhat$. Similarly, $D_{jr}^{\text{la}^2}$ tends to
$D_{1j}^{\text{la}^2}$ at large $|\yhat|$.

At all orders, since this correction affects just the first two
emissions, we only need to account for subsidiary
\DL radiation from the second gluon. That is, the \NNDL correction when
$k_1$ is emitted from the $(ab)$ dipole is
\begin{equation}\label{eq:nndl-la2}
  \delta N^{(\NNDL)}_{\mathcal{C},(ab),\text{la}^2}
  =
    \left(\frac{\as}{2\pi}\right)^2\omega^{\mathcal C}_{ab} N_c D_{ab}^{\text{la}^2} 
    \int_{0}^L\!\! \dd {\ell_1} \int_{\l_1}^L\!\! \dd \l_2 \;N_{g}^{(\DL)}(L-\l_2) 
  = \frac{\as}{2\pi}\omega^{\mathcal C}_{ab}
    \;
    D_{ab}^{{\text{la}}^2}
    \;
    \frac{1}{4}
    (\coshnu-1).
  \end{equation}
The total result is obtained after summing over all possible
coloured dipoles and partonic channels as written in Eq.~\eqref{eq:ndl-la2-general}.
We remind the reader that the kinematic weights
$\omega^\mathcal{C}_{ab}$ for inclusive jet ($Z$+jet) events are given in
appendix~\ref{app:dijets} (table~\ref{tab:Born-Zj}).

\subsubsection{A large-angle emission and a hard-collinear emission}
\label{sec:nndl-la-hc}

This corresponds to the case of a soft-and-large-angle gluon emission
together with a subsidiary hard-collinear emission (either a
$g \to gg$ or a $g\to q\bar q$ splitting) as displayed in
figure~\ref{fig:nndl-diagram-LAxhc}. As in the above section we split
the contribution with respect to the jet flavour
\begin{align}
  \label{eq:ndl-laxhc-general}
  h_{3,\text{la}\times\text{hc}}(\xi)
  &= f_q h_{3,\text{la}\times\text{hc}}^{(q)}(\xi) + f_g h_{3, \text{la}\times\text{hc}}^{(g)}(\xi)
  \quad
  \text{with}\\
  \as h_{3,\text{la}\times\text{hc}}^{(i)}(\xi) 
  &= \sum_{\mathcal{C}}
  \frac{\sigma_{\mathcal{C}}^{(i)}}{\sigma^{(i)}}
  \sum_{(ab)\in \text{event}}  \delta N^{(\NNDL)}_{\mathcal{C},(ab),\text{la}\times\text{hc}}.\nonumber
\end{align}
The starting point for computing $\delta N^{(\NNDL)}_{\mathcal{C},(ab),\text{la}\times\text{hc}}$ is
the result of section~\ref{sec:ndl-pp}, where we calculated the correction
to the Lund multiplicity due to a primary, large-angle emission. 
Instead of dressing this soft, large-angle, gluon with a tower of \DL
emissions, we now have to account for a single hard-collinear
correction anywhere in this chain of gluons.
For this, we can recycle the \NDL hard-collinear correction to the
multiplicity, $\delta N_{g,\text{hc}}^{(\NDL)}(L;\ell)$, computed
between the relative transverse momentum scale $e^{-\ell}p_\perp R$ at
which the first large-angle gluon is emitted and the cut-off scale
$\ktcut=e^{-L}p_\perp R$ (see section~3.2.2 of
Ref.~\cite{Medves:2022ccw} for explicit expressions).
We therefore have
\begin{align}
\label{eq:nndl-la-times-hc}
  \delta N^{(\NNDL)}_{\mathcal{C},(ab),\text{la}\times\text{hc}}
  &=\frac{\alpha_s}{\pi} \omega^{\mathcal C}_{ab}D_{ab}^{\text{la}}\int_0^L d\ell
  \,\delta N_{g,\text{hc}}^{(\NDL)}(L;\ell) \\
  &= \frac{\alpha_s}{2\pi}\omega^{\mathcal C}_{ab}D_{ab}^{\text{la}}
  \left[(B_{gg}+\cdiff B_{gq})\nu\sinhnu
    +2 B_{gq}(1-\cdiff)(\coshnu-1)\right], \nonumber
\end{align}
where $\cdiff = (2C_F - C_A)/C_A$.
The total result is obtained after summing over all possible coloured
dipoles and partonic channels as written in
Eq.~\eqref{eq:ndl-laxhc-general}.  Again, the kinematic weights
$\omega^{\mathcal C}_{ab}$ can be found in table~\ref{tab:Born-Zj} for
$Z$+jet events and in appendix~\ref{app:dijets} for an inclusive jet
sample.
%
\subsubsection{A large-angle emission and a running-coupling correction}
\label{sec:nndl-la-rc}

The structure of this correction is similar to the one in the previous
section with the main difference being that the running-coupling correction
can affect either the soft large-angle emission itself, or any
subsidiary soft-and-collinear emission.
We therefore write
\begin{align}
\label{eq:nndl-la-times-rc}
\delta N^{(\NNDL)}_{\mathcal{C},(ab),\text{la}\times\text{rc}}
  & = \frac{\alpha_s}{\pi} \omega^{\mathcal C}_{ab} D_{ab}^{\text{la}} \int_0^L d\ell
  \left[ 2 \as \beta_0 \ell N_g^{(\DL)}(L-\ell)
  + \delta N_{g,\text{rc}}^{(\NDL)}(L;\ell)\right] \\ \nonumber
 &  = \frac{\as}{2C_A} \omega^{\mathcal C}_{ab} D_{ab}^{\text{la}} \frac{\beta_0}{2}
  \nu \left[\nu\coshnu+\sinhnu \right] ,
\end{align}
where $\delta N_{g,\text{rc}}^{(\NDL)}(L;\ell)$ is the \NDL resummed
running-coupling correction of a gluon radiating from scale
$k_t=p_\perp R e^{-\ell}$ to $\ktcut$ (see Eq.~(4.60) of
Ref.~\cite{Medves:2022ccw} for an explicit expression).
The first (second) term in the integrand corresponds to
dressing the large-angle emission (any subsidiary emission) with a
running-coupling correction, respectively.
As before, the full \NNDL correction
$h_{3, \text{la}\times\text{rc}}(\xi) $ is obtained by summing the
contributions in Eq.~\eqref{eq:nndl-la-times-rc} for each dipole
and partonic channel, as in Eq.~\eqref{eq:ndl-laxhc-general}.

\subsection{Final result}
For completeness, we summarise here the necessary expressions needed
to compute the average Lund multiplicity within a high-energy anti-$k_t$ jet 
up to \NNDL accuracy, i.e.\ up to $h_3$ in Eq.~\eqref{eq:log-counting}. 
We define $\as
\equiv \as(p_\perp R), L \equiv \ln(p_\perp R/\ktcut), \xi \equiv
\as L^{2}, \nu \equiv \sqrt{2C_A \xi/\pi }, \cdiff \equiv
(2C_F-C_A)/C_A$ and $f_{q,g} \equiv \sigma_{q,g} / (\sigma_q + \sigma_g)$
the Born-level fractions of quark and gluon jets.
The \DL function is
\begin{align}
  \label{eq:h1-pp}
  h_1(\xi) = f_q h_{1,\text{coll}}^{(q)}(\xi) + f_g h_{1,\text{coll}}^{(g)}(\xi)
  \qquad\text{ with }\quad
  h_{1,\text{coll}}^{(i)}(\xi) = 1+\frac{C_i}{C_A}\left(\coshnu-1\right).
\end{align}
The \NDL function is
\begin{align}
\label{eq:h2-pp}
h_2(\xi)
  =  f_q h_{2, \text{coll}}^{(q)}(\xi) +
    f_g h_{2, \text{coll}}^{(g)}(\xi)
	+ \frac{1}{\sqrt {2\pi C_A}} \sum_{i \in (q,g)} \sum_{\substack{\mathcal{C}\in\text{flavour}\\\text{channels}}} \frac{\sigma_{\mathcal{C}}^{(i)}}{\sigma^{(i)}} 
	 \sum_{(ab)\in \text{event}} \omega_{ab}^\mathcal{C} D_{ab}^{\text{la}}\sinh \nu,
\end{align}
and the \NNDL function is
\begin{align}
  \label{eq:h3-pp}
  2\pi h_3(\xi)
  &=
    2\pi[f_q h_{3, \text{coll}}^{(q)}(\xi) +
    f_g  h_{3, \text{coll}}^{(g)}(\xi)]
  +
    D^{q}_{\text{hme}} \left[
    \frac{C_F}{C_A}(\cosh{\nu}-1) + 1 \right]
    + D^{g}_{\text{hme}} \cosh{\nu}\\
   & +\sum_{i \in (q,g)} \sum_{\substack{\mathcal{C}\in\text{flavour}\\\text{channels}}} \frac{\sigma_{\mathcal{C}}^{(i)}}{\sigma^{(i)}}
	 \sum_{(ab)\in \text{event}} \omega^{\mathcal C}_{ab}
   \bigg\{
    D_{ab}^{{\text{la}}^2}
    \;
    \frac{1}{4}
  (\coshnu-1)
     + D^{\text{la}}_{ab}  \frac{\pi\beta_0}{2C_A}\;
    \nu [\nu\coshnu + \sinhnu]  \nonumber\\
  & \hphantom{\sum_{i \in (q,g)} \sum_{\substack{\mathcal{C}\in\text{flavour}\\\text{channels}}} \frac{\sigma_{\mathcal{C}}^{(i)}}{\sigma^{(i)}}
  \sum_{(ab)\in \text{event}} \omega^{\mathcal C}_{ab}}
  + D^{\text{la}}_{ab}
    [(B_{gg} + \cdiff B_{gq})\nu \sinhnu +
    2B_{qg}(1-\cdiff)(\coshnu-1)] \bigg\}. \nonumber
\end{align}
In the above expressions, we have recycled the
collinear pieces of the $e^+e^-$ results, i.e.\ $h_{2, \text{coll}}^{(i)}(\xi)$ 
is defined Eq.~\eqref{eq:h2-coll}, and $h_{3, \text{coll}}^{(q)}(\xi)$ and $h_{3,
\text{coll}}^{(g)}(\xi)$ are defined in Eq.~\eqref{eq:h3-coll-q}
and~\eqref{eq:h3-coll-g}, respectively.  
%

\subsection{From fixed kinematics to fiducial cuts}\label{sec:integrated-born-phase-space}
So far, we have considered contributions to the
average Lund multiplicity stemming from jets with a fixed
transverse momentum and rapidity. In order to provide
realistic predictions for a collider environment, we have to 
relax that assumption and average our analytic result with the 
cross-section for producing jets within a given acceptance, i.e. 
\begin{equation}\label{eq:N-convolution}
  \avnlpbar(\as; \ln(p_{\perp,\text{min}}R/\ktcut)) =
  \frac{1}{\sigma} \int_{p_{\perp,\text{min}}} \frac{\dd^2 \sigma}{\dd p_\perp \dd y} \,
  \langle N^{\text{(Lund)}}(\as; L_\text{jet})\rangle \;\dd p_\perp\dd y,
\end{equation}
where $p_{\perp,\text{min}}$ is the minimum jet transverse momentum
selected, $\dd^2 \sigma/\dd p_\perp \dd y$ is the double-differential
distribution in both the transverse momentum $p_\perp$ and rapidity
$y$ of the jet.
Finally, $\sigma$ is the total jet cross-section, meaning that we are
effectively computing the average multiplicity per jet passing the
cuts.
At \NNDL accuracy, our expressions for the multiplicity already
include, through the contribution from the top of the Lund plane
derived in section~\ref{sec:nndl-hard-me}, the non-trivial NLO
corrections from the hard matrix element. The convolution in
Eq.~(\ref{eq:N-convolution}) should therefore be performed using the
Born-level cross-section. 

In practice, one could compute the average multiplicity in jets as a
function of $p_\perp$ and $y$, and perform the
integration in Eq.~(\ref{eq:N-convolution}) explicitly.
However, this integration can be simplified by realising that, 
due to the steeply falling nature of the jet spectrum, 
$p_\perp$ and $p_{\perp,\text{min}}$ are commensurate.
One can thus replace $L_\text{jet}$ by
$L\equiv L_\text{min}=\ln(p_{\perp,\text{min}}R/\ktcut)$ as the
resummation scale in Eq.~(\ref{eq:log-counting}). Since
$L_\text{jet}-L\equiv \delta_L=\ln(p_\perp/p_{\perp,\text{min}})$ does
not bring new large logarithms, one can expand the right hand side of
Eq.~\eqref{eq:N-convolution}, up to \NNDL accuracy, as
\begin{align}\label{eq:expressions-xR-xL}
  \langle N^{\text{(Lund)}}(\as;L_\text{jet})\rangle 
   & = h_1(\xi)
    + \sqrt{\as} \big[h_2(\xi) + 2 \delta_L \sqrt{\xi}\, h_1^\prime(\xi)\big]\\
  &+ \as \big[h_3(\xi)
    + (\delta_L^2 - 2 \beta_0 \delta_L\, \xi) h_1^\prime(\xi)
    + 2 \delta_L^2\, \xi\, h_1^{\prime\prime}(\xi)
    + 2  \delta_L \sqrt{\xi}\, h_2^\prime(\xi)\big],\nonumber
\end{align}
with $\xi=\as L^2$, $\as = \as(\ptmin R)$ and $h_i^\prime(\xi)$ ($h_i^{\prime\prime}(\xi)$) the first
(second) derivative of $h_i(\xi)$ with respect to $\xi$.
The main advantage of this approach is that the resummation functions
$h_i(\xi)$, given by Eqs.\eqref{eq:h1-pp}-\eqref{eq:h3-pp} are
independent of the jet $p_\perp$ and $y$.
The integration over the jet spectrum in Eq.~(\ref{eq:N-convolution}) 
thus reduces to computing a few specific quantities such as 
\begin{subequations}\label{eq:averagedL-def}
  \begin{align}
    \bar \delta_L
    &= \frac{1}{\sigma} \int \frac{\dd^2\sigma}{\dd p_\perp\dd y}
      \ln \frac{p_\perp}{\ptmin} \; \dd p_\perp\dd y,
      \label{eq:avg_deltaL}\\
    \bar \delta_L^2
    &= \frac{1}{\sigma} \int \frac{\dd^2\sigma}{\dd p_\perp\dd  y}
      \ln^2\frac{p_\perp}{\ptmin} \; \dd p_\perp\dd y,
      \label{eq:avg_deltaL2}
  \end{align}
\end{subequations}
as well as quantities related to the the radiation of soft gluons at large angles
entering at \NDL and \NNDL, i.e.\footnote{In practice, one needs to
  integrate over the full kinematics of the $2\to 2$ Born event since
  the $\omega_{ab}^{\mathcal{C}} D_{ab}^{\text{la}}$ and
  $\omega_{ab}^{\mathcal{C}} D_{ab}^{\text{la}^2}$ coefficients depend
  on the relative $\hat{y}$ between the measured jet and the
  recoiling Born-level parton.}
\begin{subequations}\label{eq:averageD-def}
  \begin{align}
    \overline{\omega D^{\text{la}}_i} 
    &= \sum_{\mathcal{C}\in\text{flavour channels}}
      \frac{1}{\sigma^{(i)}} \int \frac{\dd^2\sigma^{(i)}_{\mathcal{C}}}{\dd p_\perp\dd y}
       \sum_{(ab)\in \text{event}} \omega_{ab}^{\mathcal{C}} D_{ab}^{\text{la}}
      \; \dd p_\perp\dd y,
      \label{eq:avg_Dla}\\
    \overline{\omega D^{\text{la}^2}_i} 
    &= \sum_{\mathcal{C}\in\text{flavour channels}}
      \frac{1}{\sigma^{(i)}} \int \frac{\dd^2\sigma^{(i)}_{\mathcal{C}}}{\dd p_\perp\dd y}
       \sum_{(ab)\in \text{event}}  \omega_{ab}^{\mathcal{C}} D_{ab}^{\text{la}^2}
      \; \dd p_\perp\dd y.
      \label{eq:avg_Dla2}
  \end{align}
\end{subequations}
In Eqs.~(\ref{eq:averagedL-def}) and~(\ref{eq:averageD-def}), we can
again perform the integration using the Born-level matrix element.
All the coefficients entering Eq.~\eqref{eq:expressions-xR-xL} depend
on the underlying hard process as well as the imposed fiducial
cuts.
%
\section{Numerical results}
\label{sec:numerics}

In this section we discuss some phenomenological considerations for
the average Lund jet multiplicity, including the matching of our resummed
predictions to exact fixed-order results, as well as non-perturbative
corrections.
For definiteness, we specialise our discussion to the case of
$Z$+jet events and an inclusive jet sample in proton-proton
collisions at a centre-of-mass energy $\sqrt{s} = 13.6\TeV$.
In the $Z$+jet case we impose that $p_{\perp Z}>\ptmin=500$~GeV and $|y_Z|<2.5$
(for simplicity, we have disregarded the $Z$ decay in the fixed-order
simulations). If the event passes these cuts, we then calculate the
Lund multiplicity within the leading $R=0.8$ anti-$k_t$ jet, as
obtained from FastJet~\cite{Cacciari:2011ma}, with
$p_{\perp,\text{jet}} > p_{\perp Z}/2$
and $|y_\text{jet}| < 1.7$.
In the inclusive jet sample, we select all $R=0.8$
anti-$k_t$ jets with $p_{\perp,\text{jet}}>\ptmin= 750 \GeV$ and $|y_\text{jet}| < 1.7$.

As discussed in section~\ref{sec:integrated-born-phase-space}, the
\NNDL-accurate resummed predictions involve a series of coefficients
which depend on the underlying Born-level process.
We obtain the $Z$+jet ones from
{\tt{MadGraph}}~\cite{Alwall:2014hca} and the inclusive jet ones
from a version of
{\tt{NLOJet++}}~\cite{Nagy:2003tz} augmented so as to provide parton
flavour information~\cite{Banfi:2007gu}.
All the results reported here use the central PDF of the PDF4LHC21
set~\cite{PDF4LHCWorkingGroup:2022cjn}.
We set the central renormalisation scale to
$p_{\perp,\text{min}}R$ and the central factorisation scale to
$\hat{H}_T$, the scalar sum of the transverse momentum of the
final-state partons.
The values for the resummation coefficients with this
set of choices are listed in
table~\ref{table:multiplicity-coefficients}.

\subsection{Resummation}
\label{sec:num-resum}

\begin{table}
  \renewcommand{\arraystretch}{1.3}
  
  \begin{center}
    \begin{tabular}{l c c}
      \toprule
      & \multicolumn{2}{c}{Coefficient}\\
      & Incl.\ jets & $Z$+jet\\
      \midrule
      kinematic cuts & $p_{\perp,\text{jet}}>750$~GeV
           & $p_{\perp
             Z}>500$~GeV,$p_{\perp,\text{jet}}>\frac{1}{2}p_{\perp Z}$\\
      &  ($p_{\perp,\text{min}}=750$~GeV)
      &  ($p_{\perp,\text{min}}=500$~GeV) \\
      & $|y_\text{jet}|<1.7$ 
      & $|y_Z|<2.5$, $|y_\text{jet}|<1.7$ \\
      strong coupling
      & $\as(Rp_{\perp,\text{min}}) = 0.091$
      & $\as(Rp_{\perp,\text{min}}) = 0.097$\\
      quark, gluon jet fractions
      & $f_q = 0.553$ & $f_q = 0.818$ \\
      & $f_g = 0.447$ & $f_g = 0.182$ \\
      spectrum convolutions
      & $\bar\delta_L = -0.166$ & $\bar\delta_L = -0.195$\\
      & $\bar\delta_L^2 = 0.052$ & $\bar\delta_L^2 = 0.071$\\
      hard matrix-element
      & $D_{\text{hme}}^{q} = -9.331$ & $D_{\text{hme}}^{q} = -11.545$ \\
      & $D_{\text{hme}}^{g} = 5.508$ & $D_{\text{hme}}^{g} = 8.285$\\
      two large-angle emissions
      & $\overline{\omega D^{\text{la}^2}_q} = 0.030$
      & $\overline{\omega D^{\text{la}^2}_q} = -0.074$\\
      & $\overline{\omega D^{\text{la}^2}_g} = -0.246$
      & $\overline{\omega D^{\text{la}^2}_g} = -0.504$ \\
      \bottomrule
    \end{tabular}
  \end{center}
  \caption{Coefficients entering the \NDL $h_2$ and \NNDL $h_3$ functions 
  for our two jet samples, inclusive jets and $Z$+jet events with
  $R=0.8$ in both cases.}\label{table:multiplicity-coefficients}
\end{table}

\begin{figure}
  \begin{subfigure}[t]{0.48\linewidth}
    \includegraphics[scale=0.73,page=1]{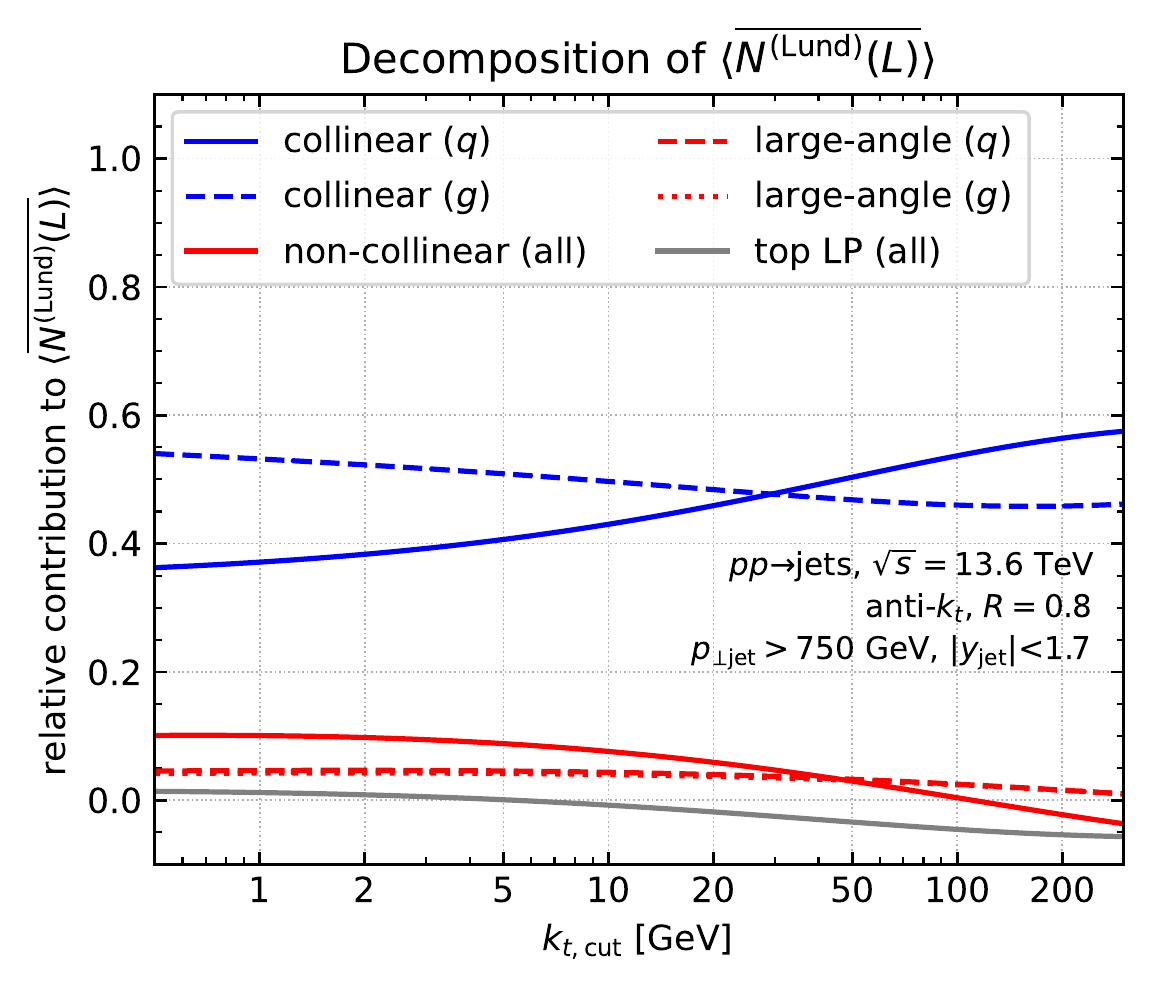}%
    \caption{}\label{fig:resum-coll-vs-la-jj}
  \end{subfigure}
  \hfill%
  \begin{subfigure}[t]{0.48\linewidth}
    \includegraphics[scale=0.73,page=2]{Plots/plot-q-g-rest.pdf}%
    \caption{}\label{fig:rel-resum-coll-vs-la-zj}
  \end{subfigure}
  \caption{Relative weight of each term to the total Lund multiplicity 
    at \NNDL accuracy for an inclusive sample of jets (left)
    and $Z$+jet events (right). We distinguish universal, collinear terms 
    (blue) from those contributions
    which depend on the event structure, i.e.\ wide-angle radiation (red) 
    and hard matrix-element corrections (grey).
    }
  \label{fig:coll-vs-la-plot}
\end{figure}

Throughout this paper, we have organised the calculation of the
average Lund multiplicity as the sum of a collinear piece, identical
to the $e^+e^-$ result, and a large-angle piece.
The collinear contribution is universal in the sense that it is
process-independent and  only depends on the Born-level jet flavour.
Conversely, the large-angle contribution depends on the full structure
of the hard process.
Up to \NNDL accuracy, it can be further split into a contribution
associated with soft emissions (close to the edge of the jet), and
genuine fixed-order corrections associated with an emission at the top
of the Lund plane. The former is still associated with the Born-level
jet flavour, but for the latter the breakdown into a quark and a gluon
contribution is ambiguous.

To gain insight into the relative sizes of each of these
  contributions we plot in figure~\ref{fig:coll-vs-la-plot} the relative
weights of the different contributions for both the inclusive jet
sample and $Z$+jet events.
Both the soft and the collinear contributions have explicitly been split
according to flavour, and the contribution labelled as
``non-collinear'' includes both the contributions from soft-large-angle
radiation as well as the contributions from the top of the Lund plane. 
For both hard processes, almost $90\%$ of the Lund multiplicity is 
dominated by contributions of collinear origin.
Focusing on the large-angle corrections, one sees that the
contribution from hard emissions, i.e.\ at the top of the Lund plane,
is further suppressed (roughly by a factor 2) compared to the soft-and-wide-angle 
contributions. The hard contribution is however larger for the
$Z$+jet case than for inclusive jets, and, in both cases, changes sign
as a function of $\ktcut$.
Globally speaking, this picture is in agreement with the fact that
collinear effect start at \DL accuracy, soft wide-angle effects start
at \NDL and top-of-the-Lund-plane corrections only contribute at
\NNDL.
The fact that the number of emissions inside a jet is dominated by
universal collinear physics depending only on the Born-level 
flavour of the jet is in quantitative agreement with the study 
of the dependence of quark and gluon jets on the topology of 
the hard process from Ref.~\cite{Bright-Thonney:2018mxq}.

Regarding the flavour decomposition of the result, the relative sizes
of quark and gluon contributions differ drastically between the
inclusive and the $Z$+jet sample.
As expected, we find a larger relative contribution of quark-initiated
jets in $Z$+jet events than in the inclusive sample. In the latter,
the slightly higher initial fraction of quark-initiated jets (see table~\ref{table:multiplicity-coefficients}) is
compensated by Casimir scaling for small values of $\ktcut$, 
i.e.\ gluons give a larger multiplicity than quarks at small $\ktcut$. However,
quark jets dominate down to $\ktcut \sim 30\GeV$.
%

\subsection{Phenomenological predictions}
\label{sec:pheno}
To produce realistic predictions for phenomenological applications at
the LHC we match our resummed calculation, computed in
section~\ref{sec:pp-calc}, to the exact fixed-order result at NLO,
i.e.\ $\order{\as^2}$ compared to the underlying $2\to 2$ hard
process.\footnote{The first non-trivial order for the Lund
  multiplicity requires at least one emission on top of the
  underlying $2\to 2$ process. This means that the LO and NLO
  fixed-order multiplicities we refer to in this section correspond
  respectively to a relative $\order{\as}$ and $\order{\as^2}$
  compared to the hard $2\to 2$ process.}
We further account for theoretical uncertainties and include
non-perturbative corrections.
We perform these three steps in this section, focusing only on our
inclusive jet sample for simplicity.

\paragraph{Theoretical uncertainties.}
We evaluate the uncertainties associated with missing higher orders by
varying the renormalisation ($\ptmin R$), factorisation ($\hat{H}_T$)
and resummation scales ($\ptmin R/\ktcut$).
To this end we multiply each of the scales by a dimensionless
parameter, $x_R$, $x_F$ and $x_L$ respectively. Variations of $x_F$
only affect quantities that depend on the averaging over the
Born-level jet spectrum, i.e.\ quantities listed in
table~\ref{table:multiplicity-coefficients}. Conversely, when varying
$x_R$ and $x_L$ one introduces spurious terms that spoil the
perturbative accuracy of the resummation. To maintain \NNDL accuracy,
we introduce counter-terms so that our resummed expression becomes
\begin{align}\label{eq:N-xR-xL}
  \avnlpbar_\text{resum}
  & = \bar h_1(\xi)
    + \sqrt{\as} \big[h_2(\xi)
    + 2 (\bar\delta_L - \ln x_L) \sqrt{\xi}\, \bar h^\prime_1(\xi)\big]\\
  &+ \as \Big\{\bar h_3(\xi)
    + \left[2 \beta_0 (\ln x_R-\bar \delta_L)\, \xi 
    + \ln^2 x_L -2\ln x_L \bar \delta_L + \bar\delta^2_L\right] \bar h_1^{\prime}(\xi)\nonumber \\
  &+ 2 (\ln^2 x_L -2\ln x_L \bar \delta_L + \bar\delta^2_L) \xi\, \bar h_1^{\prime\prime}(\xi)
    + 2 \sqrt{\xi} \Big[\overline{\delta_L h_2^\prime(\xi) }
    - \ln x_L \, \bar h_2^\prime(\xi)\Big]\Big\},\nonumber
\end{align}
where the bar indicates an average over the jet
spectrum as explained in
section~\ref{sec:integrated-born-phase-space}.
Note that, in the last term, we have used the short-hand notation
$\overline{\delta_L h_2^\prime(\xi)}$ to emphasise that, in the
large-angle \NDL contribution, one has to compute explicitly the
overall $\overline{\delta_L \omega_\mathcal{C}D_i^\text{la}}$
coefficient averaged
over Born-level events. All the other, collinear, \NDL terms simply
involve $\bar\delta_L$.
In practice, we vary separately $x_R$, $x_F$ and $x_L$ between 0.5 and 2. 
Our total perturbative uncertainty is obtained by summing in
quadrature the 7-point variation~\cite{Cacciari:2003fi} of $x_R$ and
$x_F$ with the resummation uncertainty obtained by varying $x_L$.

\begin{figure}
  \begin{subfigure}[t]{0.48\linewidth}
    \includegraphics[scale=0.73,page=1]{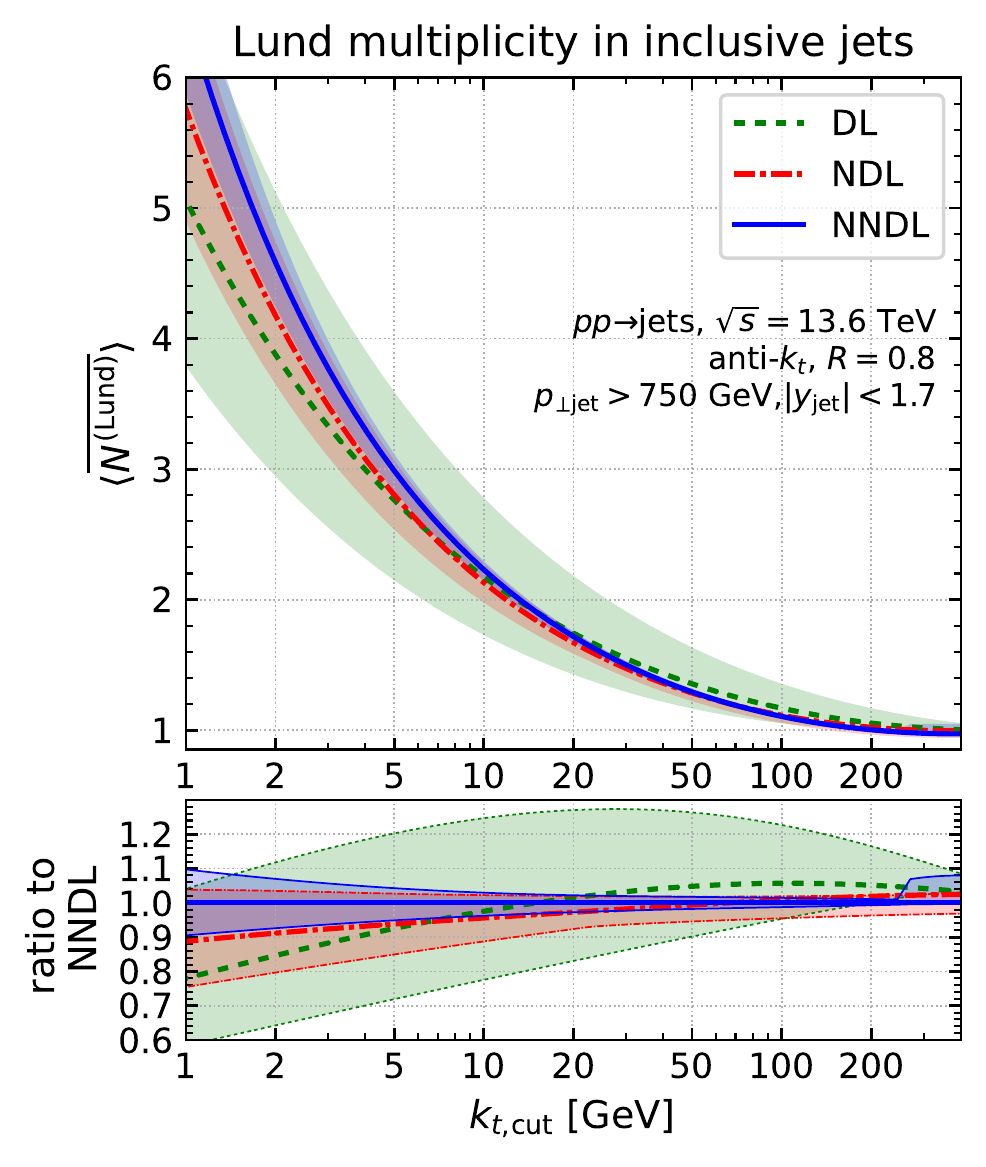}%
    \caption{}\label{fig:lhc-plot-resum}
  \end{subfigure}
  \hfill%
  \begin{subfigure}[t]{0.48\linewidth}
    \includegraphics[scale=0.73,page=2]{Plots/plot-match-jj-R0.8.pdf}%
    \caption{}\label{fig:lhc-plot-match}
  \end{subfigure}
  \caption{Left: Resummed Lund average multiplicity at \DL (dashed, green),
    \NDL (dash-dotted, red) and \NNDL (solid, blue) as a function of
    $\ktcut$ for LHC kinematics.
    Right: Average Lund multiplicity after matching with NLO. The pure NLO
    result is shown in dotted (grey) for reference, together with
    matched NLO+\NDL (dash-dotted, red) and NLO+\NNDL (solid, blue).
    The bottom panels display the ratio to the most
    accurate result (\NNDL on the left and NLO+\NNDL on the right).
  }
  \label{fig:lhc-plot}
\end{figure}
The numerical evaluation of Eq.~\eqref{eq:N-xR-xL} 
is displayed in figure~\ref{fig:lhc-plot-resum} at \DL, \NDL and \NNDL
accuracy. 
Let us first focus on the central values.  We observe that the
\NDL correction enhances the Lund multiplicity with respect to the \DL
result when $\ktcut\lesssim 6\GeV$. We note that this was not the case
in the $e^+e^-$ result, where the \NDL correction reduced the
multiplicity for all $\ktcut$ values. This increase in the
multiplicity of high-energy jets is solely due to the large-angle
emission correction calculated in section~\ref{sec:ndl-pp}.
The \NNDL corrections further increase the Lund multiplicity
for $\ktcut \lesssim 15\GeV$, and have only moderate impact
for all larger values of $\ktcut$.
Turning to the uncertainty band, we find that 
it is remarkably reduced beyond \DL accuracy. For example, 
at $\ktcut = 5\GeV$ ($10\GeV$), the error band shrinks from a
$\sim 28\%$ ($\sim 24\%$) 
at \DL to $\sim 10\%$ ($\sim 7\%$) at \NDL accuracy and further to 
$\sim 5\%$ ($\sim 3.4\%$) at \NNDL accuracy, i.e.\ an additional
reduction around 50\%.
For $\ktcut\gtrsim\ptmin R /2=300$ GeV, well beyond 
the regime of validity of the resummed result, 
we note the irregular behaviour of
the uncertainties associated with the \NDL and \NNDL accurate
results. This unphysical behaviour disappears after 
matching as we discuss next.

\paragraph{Matching.} We use an additive scheme for the Lund multiplicity:
\begin{align}
  \label{eq:matching}
  \avnlpbar_\text{match}
  &=
  \avnlpbar_\text{fo} \\
  &+ \Big(
    \avnlpbar_\text{resum} - \avnlpbar_\text{resum,fo}
  \Big) 
  \Big(1 - \frac{2\ktcut}{\ptmin R}\Big)\Theta{\Big(\ktcut < \frac{\ptmin R}{2}\Big)},
  \nonumber
\end{align}
where $\avnlpbar_\text{fo}$ is the exact fixed-order result obtained from
\texttt{NLOJet++} taken here at NLO, i.e.\ at $\order{\as^2}$ relative to
the Born-level event, $\avnlpbar_\text{resum}$ is the \NNDL-accurate
resummed result, Eq.~\eqref{eq:N-convolution}, and we subtract
$\avnlpbar_\text{resum,fo}$, the
$\order{\as^2}$ expansion of the resummed result, 
to avoid double counting.
We also introduce a damping factor to smooth the transition between
the resummed and fixed-order results that switches off the resummation
at $\ktcut^{\text{max}} = \ptmin R/2$ where it should anyway not be
trusted.
This ensures that the matched
results reduce to the fixed-order at large values of $\ktcut$, while
not spoiling the \NNDL accuracy at small $\ktcut$ values.\footnote{One
  could alternatively have redefined $L$ so that
  $L\to \ln(\ptmin R/k_t)$ at small $k_t$ and
  $L\to 0$ when $k_t$ is large. Note also that, compared to our
  earlier $e^+e^-$ study, the $k_t$ spectrum no longer has a
  well-defined fixed-order endpoint but has instead a tail extending far
  beyond $\ptmin R$.}
We note that scale uncertainties are obtained using
Eq.~(\ref{eq:N-xR-xL}) for the resummed part. A similar expression is
used for the fixed-order expansion $\avnlpbar_\text{resum,fo}$ where
the $h_i(\xi)$ functions and their derivatives are expanded in
$\as$. In the exact fixed-order results we use the $x_R$, $x_F$ scale
variations as provided by \texttt{NLOJet++}.

We show in figure~\ref{fig:lhc-plot-match} the effect of matching our
\NDL and \NNDL accurate resummed result with the exact NLO result, as
described in Eq.~\eqref{eq:matching}. We observe that the resummation
slightly increases the value of the Lund multiplicity with an effect
more pronounced as $\ktcut$ decreases. As an example, taking the
central value of Lund multiplicity curves at $\ktcut=5\GeV$
we find that $\avnlpbar$ is $\approx 2.6$ at NLO and $2.7$
at NLO+\NNDL.
The uncertainty band after matching increases due to the impact of
$x_R$ and $x_F$ variations on the fixed-order result. Nevertheless,
the error band on the NLO+\NNDL result remains small, with a value of
$\sim 6\%$ at $5\GeV$ and below 15\% down to 1~GeV, as observed in the
bottom panel of figure~\ref{fig:lhc-plot-match}.
We also note that the matching procedure indeed washes out the
unphysical behaviour observed at large $\ktcut$ values for the
resummed predictions in figure~\ref{fig:lhc-plot-resum}.

\paragraph{Non-perturbative corrections.}
The last ingredient that we incorporate into our 
theoretical prediction alongside matching and scale variations
is non-perturbative corrections due to hadronisation 
and multi-parton interactions (MPI).
To estimate these we run the
\texttt{Pythia8.3}~\cite{Bierlich:2022pfr} (with \texttt{Monash13}~\cite{Skands:2014pea},
\texttt{4C}~\cite{Corke:2010yf}, and \texttt{Atlas14}~\cite{ATL-PHYS-PUB-2014-021} 
tunes), \texttt{Herwig7.20}~\cite{Bellm:2015jjp} and \texttt{Sherpa2.2.11}~\cite{Sherpa:2019gpd}
event generators at parton level and at hadron level with MPI turned on.

\begin{figure}
  \begin{subfigure}[t]{0.48\linewidth}
    \includegraphics[scale=0.73,page=1]{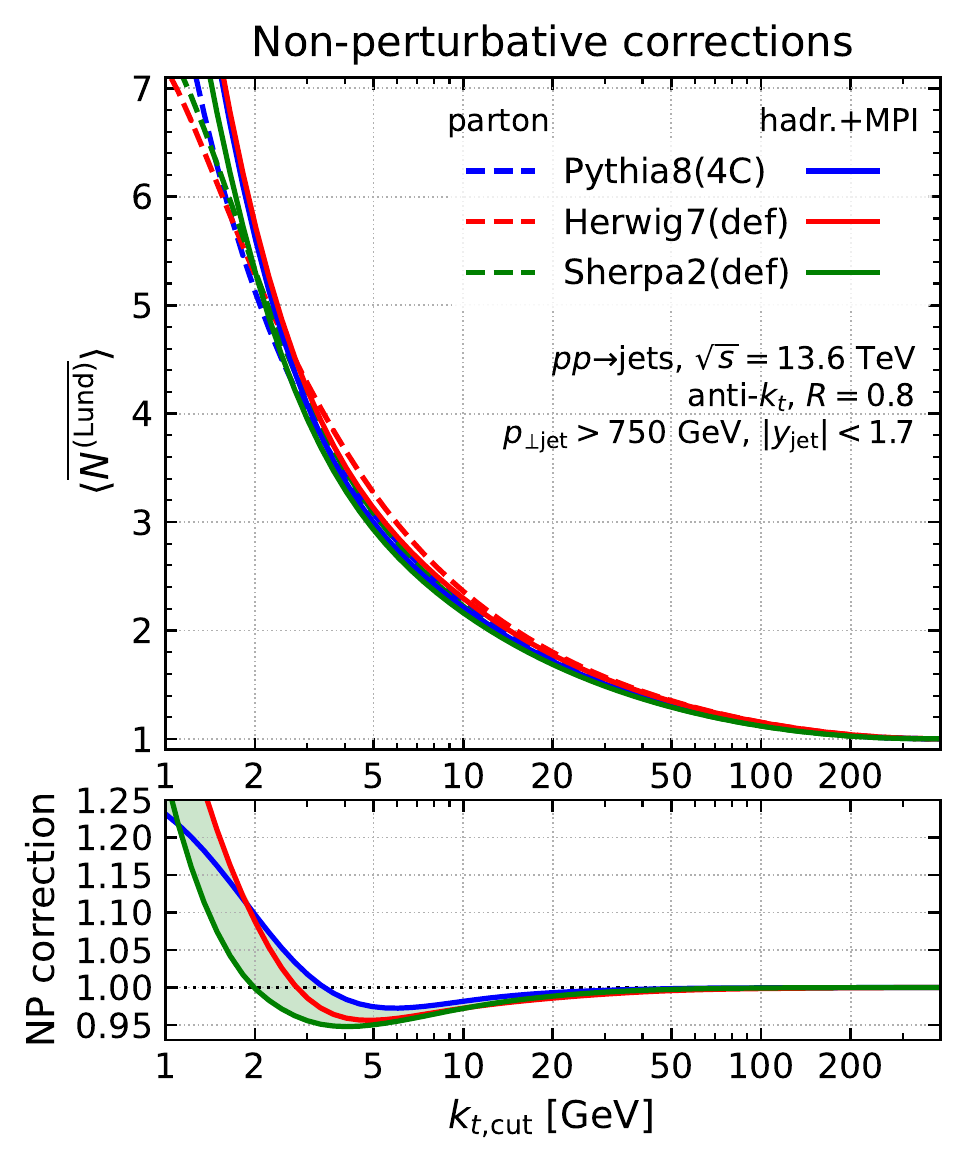}%
    \caption{}\label{fig:np-corr}
  \end{subfigure}
  \hfill%
  \begin{subfigure}[t]{0.48\linewidth}
    \includegraphics[scale=0.73,page=1]{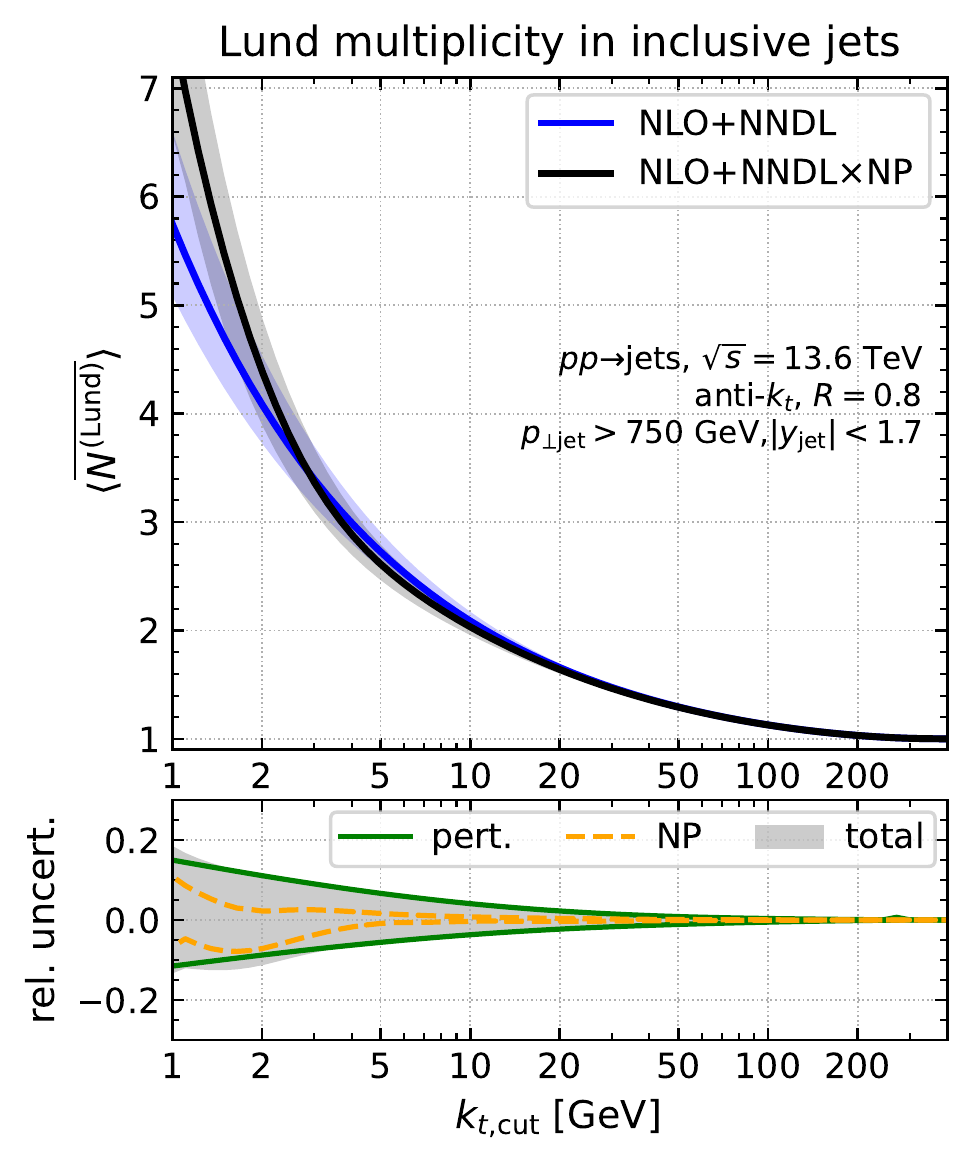}%
    \caption{}\label{fig:final-result}
  \end{subfigure}
  \caption{Left: Lund multiplicity distributions obtained from Monte
    Carlo simulations (upper panel) together with the hadron+MPI/parton ratio
    (lower panel) used as our estimate of the non-perturbative corrections.
    Results are shown for \texttt{Pythia8} (with the \texttt{4C}
    tune), \texttt{Sherpa2} and \texttt{Herwig7} (with their default
    tune).
    The (green) shaded area in the bottom panel corresponds to the envelope of the individual non-perturbative corrections.
    Right: prediction for the average Lund multiplicity in LHC
    high-energy jets. The matched \NNDL+NLO perturbative results are
    showed both with (black) and without (blue) non-perturbative corrections.
    The bottom panel displays the breakdown of uncertainties in their
    perturbative and non-perturbative
    contributions. }\label{fig:np-corr-and-final}
\end{figure}

In figure~\ref{fig:np-corr} we show the average Lund multiplicity
distributions obtained from our Monte Carlo simulations as well as the
ratio between the hadron-level (including MPI) and parton-level
results. For readability purposes, we only show the
\texttt{Pythia8} results for the \texttt{4C} tune, which
typically drives the upper edge of the non-perturbative corrections
(with the other two \texttt{Pythia8} tunes showing
intermediate non-perturbative corrections).
Overall the non-perturbative corrections are small, staying within
$\sim 5\%$ down to $\ktcut \sim 3 \GeV$.

Below $\ktcut\sim 2\GeV$ the non-perturbative corrections start
to become significant.
However, as noted in Ref.~\cite{Medves:2022ccw}, our resummation
becomes questionable for values of $L$ of order $1/(\beta_0\as)$. This
can be traced to the fact that running-coupling corrections dominate
the resummation functions $h_k(\xi)$ at large $\xi$ with contributions
proportional to $\xi^{k-1}h_1(\xi)$. In this regime, which corresponds
to $\ktcut\lesssim 4$~GeV with our kinematic cuts, one should also
consider resumming these running-coupling corrections beyond their
expansion to \NNDL.

In practice we implement the average of the hadron/parton ratios from
the five Monte Carlo setups as a multiplicative factor into the
perturbative results. The envelope of these five hadron/parton ratios
(the shaded region in the bottom panel of figure~\ref{fig:np-corr}) is
used as an estimate of the uncertainties associated to
non-perturbative effects.
When applying non-perturbative corrections to our resummed (or
NLO-matched) results, the non-perturbative uncertainties are added in
quadrature with the perturbative uncertainties.

The main phenomenological result of this paper is displayed in
figure~\ref{fig:final-result} where we show the average Lund
multiplicity for our inclusive jet sample.
We present results at \NNDL accuracy matched to NLO with (black) and
without (blue) non-perturbative corrections.
As expected from figure~\ref{fig:np-corr}, the effect of the
non-perturbative corrections starts being clearly visible for $\ktcut$
values of a few GeV and below.
The theoretical uncertainties are displayed in the lower panel of 
figure~\ref{fig:final-result}. They are dominated by the perturbative
uncertainties across the full $\ktcut$ range, with non-perturbative
corrections being negligible for $\ktcut\gtrsim 4$~GeV.
This suggests that there is room for further theoretical developments
on the perturbative part of the calculation.
This includes subleading all-order contributions (N$^{k\ge 3}$DL),
subleading fixed-order results
(N$^{k\ge 2}$LO)~\cite{Abreu:2021oya,Czakon:2021mjy} or multiple
running-coupling correction effects discussed above and relevant for
$\ktcut\lesssim 4$~GeV.

\section{Conclusions}
\label{sec:conclusions}
A new jet observable, dubbed Lund multiplicity, 
has recently been introduced in Ref.~\cite{Medves:2022ccw}. 
It provides a procedure for evaluating subjet multiplicity with the
use of an IRC-safe cutoff $\ktcut$ in the spirit of modern Lund-plane
declustering techniques~\cite{Dreyer:2018nbf}.
In Ref.~\cite{Medves:2022ccw}, the event-wide average Lund
multiplicity was resummed up to \NNDL accuracy in $e^+e^-$ collisions
and up to \NDL accuracy in colour singlet production in $pp$
collisions.
The present manuscript extends the definition and calculation of the
average Lund multiplicity to the case of high-energy jets at hadron
colliders by counting the mean number of subjets per anti-$k_t$ jet
with relative $k_t$ above $\ktcut$.
The main motivation behind this work is to provide a
multiplicity-based observable which is perturbatively well-defined in
QCD, and can benefit from the large energy range accessible at the
LHC. Indeed, for a jet of transverse momentum $p_\perp$ and radius $R$,
the Lund multiplicity probes energy scales from $\ktcut$ values around
1~GeV up to $p_\perp R$ values of several hundreds of GeV.

From a theoretical standpoint, the main achievement of this paper is
the all-order computation of the average Lund multiplicity up to \NNDL
accuracy, which is particularly relevant given the large range
of transverse momenta probed by this observable.
There are two main differences with respect to the $e^+e^-$ event-wide
result: the notion of a jet radius, which impacts the large-angle
components of the resummation starting at \NDL accuracy (see
sections~\ref{sec:ndl-pp} and \ref{sec:NNDL-accuracy}), and the
presence of experimental fiducial cuts used for the jet analysis in a
collider environment (see
section~\ref{sec:integrated-born-phase-space}).
For a jet of fixed $p_\perp$ and rapidity, \NNDL accuracy is reached by recycling
universal contributions sensitive to collinear emissions and running
coupling corrections (see section~\ref{sec:recape+e-}) from the
results in $e^+e^-$ collisions, and supplementing these with
jet-radius effects specific to the case of high-energy jets in $pp$
collisions.
The resulting \NNDL distribution for a generic $2\to 2$ hard process,
Eqs.~\eqref{eq:h1-pp}-\eqref{eq:h3-pp}, is the main theoretical result
of this paper.
Reaching \NNDL accuracy for a realistic situation with fiducial cuts
is achieved via a simple procedure (detailed in
section~\ref{sec:integrated-born-phase-space}).

We have performed phenomenological studies specialising our discussion
to $Z$+jet and inclusive jet samples at the LHC.
Our study has shown (figure~\ref{fig:coll-vs-la-plot}) that universal,
collinear, terms contribute towards $\sim90\%$ of the average Lund
multiplicity in both hard processes. Decomposing contributions in
terms of jet flavour, we have found that the $Z$+jet result is overall
governed by quark-initiated jets, while for inclusive jets both quark-
and gluon-initiated jets contribute commensurately for all values of
the transverse momentum cut $\ktcut$.

Furthermore, we have presented (section~\ref{sec:pheno}) predictions
for the average Lund multiplicity of high-energy inclusive jets at the LHC.
In order to cover the full range of $\ktcut$ values, we have matched
our \NNDL calculation to exact NLO fixed-order results.
Non-perturbative corrections have been included via a multiplicative
factor extracted from general-purpose Monte Carlo event generators.
The result of this procedure is shown in figure~\ref{fig:np-corr-and-final}.
Theoretical uncertainties remain below $15\%$ down to $\ktcut$ values
of 1~GeV and are dominated by their perturbative component.
In particular, non-perturbative uncertainties are remarkably small
($<3\%$) for $\ktcut\gtrsim 4$ GeV, with the overall
uncertainty remaining under $\sim 7\%$.
This leaves some margin for further theoretical improvements like the
inclusion of higher-order effects such as $\order{\as^3}$ matching,
N$^3$DL resummation, or joint resummations of running-coupling effects
at small $\ktcut$.

In a more generic context, experimental measurements of the number of
charged particles within a jet~\cite{ATLAS:2016vxz,CMS:2021iwu} and of
the number of subjets inside groomed jets~\cite{ATLAS:2019kwg} have
revealed sizeable discrepancies between Monte Carlo event generators.
A high-precision calculation of a multiplicity-based observable can
therefore help clarifying this situation.
Resolving these discrepancies can then result in a better
determination of the jet energy scale~\cite{ATLAS:2017bje}, a key
quantity in any jet analysis at the LHC.
Therefore, besides their purely theoretical interest, the results presented in
this paper have the potential to serve as benchmarks
to test and develop Monte Carlo event
generators and, more concretely, to pin down the logarithmic accuracy of
their parton shower in the spirit
of~\cite{Dasgupta:2020fwr,Hamilton:2020rcu,Karlberg:2021kwr,Hamilton:2021dyz,vanBeekveld:2022ukn,vanBeekveld:2022zhl,Herren:2022jej}.
%

\section*{Acknowledgements}
We are grateful to our PanScales collaborators (Melissa van Beekveld,
Mrinal Dasgupta, Fr\'ed\'eric Dreyer, Basem El-Menoufi, Silvia
Ferrario Ravasio, Keith Hamilton, Jack Helliwell, Alexander Karlberg,
Pier Monni, Gavin Salam, Ludovic Scyboz and Rob Verheyen) for
discussions and comments on this manuscript.
This work has been supported by the European Research Council (ERC) under
the European Union’s Horizon 2020 research and innovation programme
(grant agreement No.\ 788223, PanScales).
%
\appendix
%
%
\section{Kinematic weights for dijets}
\label{app:dijets}
Here we provide the kinematic weights entering into the
formulae of section~\ref{sec:ndl-pp}. We have extracted 
them from Ref.~\cite{Ellis:1986bv} and $s,t,u$ refer 
to the usual Mandelstam variables. 

\begingroup
\allowdisplaybreaks
\begin{subequations}
\begin{align}
  \omega^{gg \to gg}_{12}(s,t,u)
  & = \omega^{gg \to gg}_{34}(s,t,u) =  \omega^{gg \to gg}_{13}(t,s,u)
    = \omega^{gg \to gg}_{24}(t,s,u) =\nonumber\\
  & = \omega^{gg \to gg}_{14}(u,t,s)
    = \omega^{gg \to gg}_{23}(u,t,s)
    = C_A\left(\frac{2}{3}+\frac{H_{1d}(s,t,u)}{H_d(s,t,u)}\right),\\
  %
  \omega^{gg \to q\bar q}_{12}(s,t,u)
  & = \omega^{qg \to qg}_{24}(t,s,u) = \omega^{q\bar q\to gg}_{34}(s,t,u)
    = 2N_c\left[1 - \left(1-\frac{1}{N_c^2}\right)\frac{H_{1c}(s,t,u)}{H_c(s,t,u)}\right]\\
  \omega^{gg \to q\bar q}_{34}(s,t,u)
  & = \omega^{q\bar q\to gg}_{12}(s,t,u) = \omega^{qg \to qg}_{13}(t,s,u)
    = 2C_F\left(1-2\frac{H_{1c}(s,t,u)}{H_c(s,t,u)}\right),\\
  \omega^{gg \to q\bar q}_{13}(s,t,u)
  & =\omega^{gg \to q\bar q}_{24}(s,t,u)
    = \omega^{q\bar q\to gg}_{13}(s,t,u)
    = \omega^{q\bar q\to gg}_{24}(s,t,u) =\nonumber\\
  & = \omega^{qg \to qg}_{12}(t,s,u) = \omega^{qg \to qg}_{34}(t,s,u)
    = 2C_F\frac{H_{1c}(s,t,u)-H_{2c}(s,t,u)}{H_c(s,t,u)},\\
  \omega^{gg \to q\bar q}_{14}(s,t,u)
  & = \omega^{gg \to q\bar q}_{23}(s,t,u)
    = \omega^{q\bar q\to gg}_{14}(s,t,u)
    = \omega^{q\bar q\to gg}_{23}(s,t,u)\nonumber\\
  & = \omega^{qg \to qg}_{14}(t,s,u) = \omega^{qg \to qg}_{23}(t,s,u)
    = 2C_F\frac{H_{1c}(s,t,u)+H_{2c}(s,t,u)}{H_c(s,t,u)},\\
  %
  \omega^{qq' \to qq'}_{12}(s,t,u)
  & = \omega^{qq' \to qq'}_{34}(s,t,u)
    = \omega^{q\bar q \to q'\bar q'}_{14}(s,t,u)
    = \omega^{q\bar q \to q'\bar q'}_{23}(s,t,u) =\nonumber\\
  & = \omega^{q\bar q' \to q \bar q'}_{14}(s,t,u)
    = \omega^{q\bar q' \to q \bar q'}_{23}(s,t,u)
    = 2/N_c,\\
  \omega^{qq' \to qq'}_{13}(s,t,u)
  & = \omega^{qq' \to qq'}_{24}(s,t,u)
    = \omega^{q\bar q \to q'\bar q'}_{12}(s,t,u)
    = \omega^{q\bar q \to q'\bar q'}_{34}(s,t,u) = \nonumber\\
  & = \omega^{q\bar q' \to q \bar q'}_{13}(s,t,u)
    = \omega^{q\bar q' \to q \bar q'}_{24}(s,t,u)
    = -1/N_c,\\
  \omega^{qq' \to qq'}_{14}(s,t,u)
  & = \omega^{qq' \to qq'}_{23}(s,t,u)
    = \omega^{q\bar q \to q'\bar q'}_{13}(s,t,u)
    = \omega^{q\bar q \to q'\bar q'}_{24}(s,t,u) = \nonumber\\
  & = \omega^{q\bar q' \to q \bar q'}_{12}(s,t,u)
    = \omega^{q\bar q' \to q \bar q'}_{34}(s,t,u)
    = 2C_F-1/N_c,\\
  %
  \omega^{qq \to qq}_{12}(s,t,u)
  & = \omega^{qq \to qq}_{34}(s,t,u)
    = \omega^{q\bar q \to q \bar q}_{14}(u,t,s)
    = \omega^{q\bar q \to q \bar q}_{23}(u,t,s) = \nonumber\\
  & = \frac{2}{N_c^2}\frac{N_cH_a(s,t,u) + N_cH_a(s,u,t) - (N_c^2+1)H_{1b}(s,t,u)}{H_b(s,t,u)},\\
  \omega^{qq \to qq}_{13}(s,t,u)
  & = \omega^{qq \to qq}_{24}(s,t,u)
    = \omega^{q\bar q \to q \bar q}_{13}(u,t,s)
    = \omega^{q\bar q \to q \bar q}_{24}(u,t,s) = \nonumber\\
  & = \frac{1}{N_c^2}\frac{-N_cH_a(s,t,u) + N_c(N_c^2-2)H_a(s,u,t) + 2H_{1b}(s,t,u)}{H_b(s,t,u)},\\
  \omega^{qq \to qq}_{14}(s,t,u)
  & = \omega^{qq \to qq}_{23}(s,t,u)
    = \omega^{q\bar q \to q \bar q}_{12}(u,t,s)
    = \omega^{q\bar q \to q \bar q}_{34}(u,t,s) = \nonumber\\
  & = \frac{1}{N_c^2}\frac{N_c(N_c^2-2)H_a(s,t,u) - N_cH_a(s,u,t) + 2H_{1b}(s,t,u)}{H_b(s,t,u)}.
\end{align}
\end{subequations}
\endgroup
In the previous equations, we have defined:
\begin{subequations}
\begin{align}
H_a &= \frac{s^2+u^2}{t^2}, &
H_b &= \frac{s^2+u^2}{t^2} + \frac{s^2+t^2}{u^2} -\frac{2}{N_c}\frac{s^2}{t u}, \\ 
H_{c} & = (t^2+u^2)\Big[\Big(1-\frac{1}{N_c^2}\Big) \frac{1}{tu} -\frac{2}{s^2}\Big], &
H_{1b} & = \frac{s^2}{tu}, \\
H_{1c}& =\frac{N_c}{4C_F}(t^2+u^2)\Big[\Big(1-\frac{2}{N_c^2}\Big)\frac{1}{tu} - \frac{2}{s^2}\Big], &
H_d & = 3 - \frac{ut}{s^2} - \frac{us}{t^2} - \frac{st}{u^2}, \\
H_{2c} & =  \frac{N_c}{4C_F}(t^2-u^2)\Big[\frac{1}{tu} - \frac{2}{s^2}\Big], &
H_{1d} & = \frac{1}{6} \Big(\frac{st}{u^2} + \frac{su}{t^2} - \frac{2tu}{s^2} - \frac{3 s^2}{tu} + 3\Big).
\end{align}
\end{subequations}

\section{Extraction of the top-of-the-Lund-plane coefficients
  $D_{\text{hme},R}^{(bc)}$}\label{app:D-hme-bc-extraction}

\begin{figure}
  \begin{subfigure}{0.48\textwidth}
    \includegraphics[width=\textwidth,page=1]{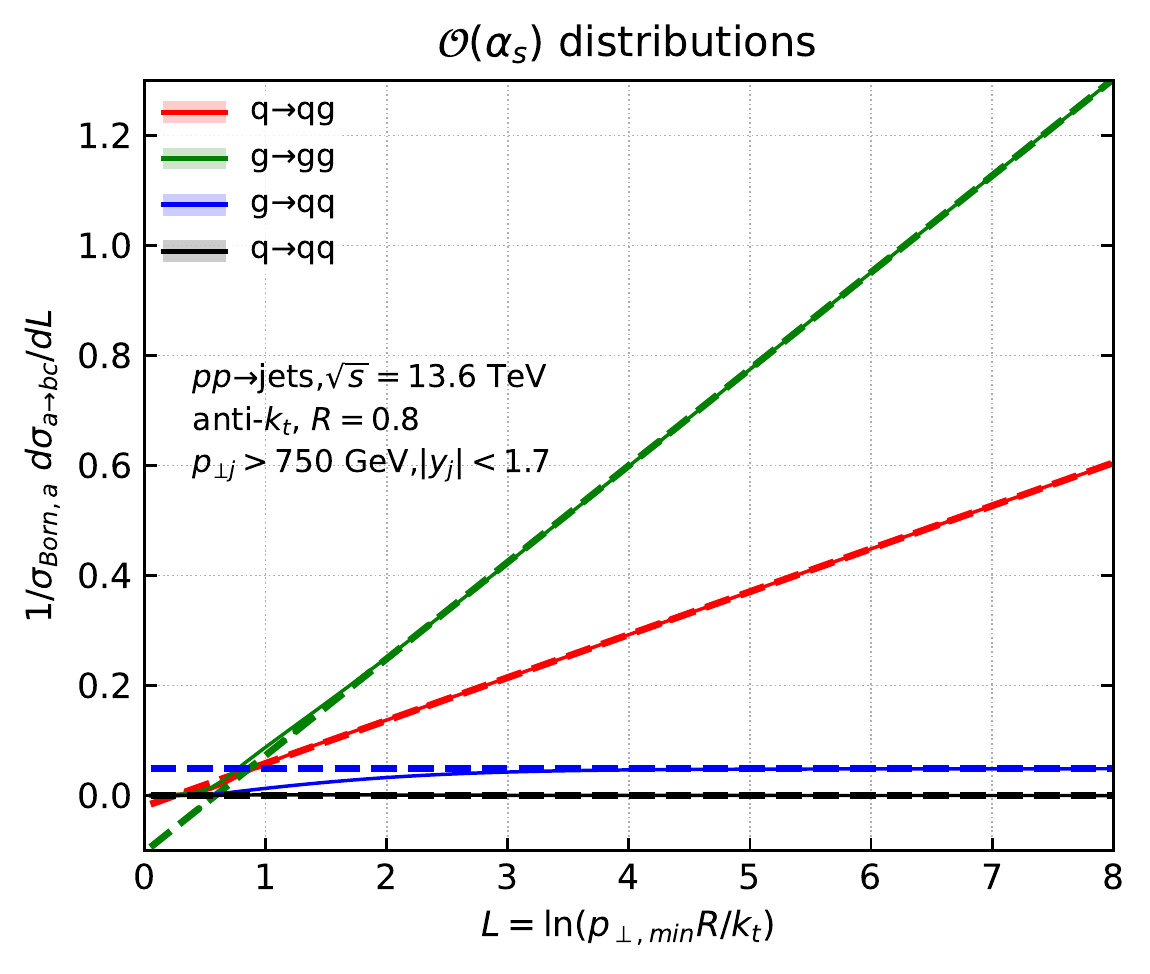}
    \caption{Differential multiplicity distributions, as obtained from
      \texttt{NLOJet++} for the real NLO correction to our inclusive
      jet sample (solid lines). The dashed lines are the \NDL
      expectations. The results are split in flavour
      channels.}\label{fig:top-extraction-initial}
  \end{subfigure}
  \hfill
  \begin{subfigure}{0.48\textwidth}
    \includegraphics[width=\textwidth,page=5]{Plots/process-resum-coefs-R0.8-paper.pdf}
    \caption{Extraction of the $D_{\text{hme},R}^{(bc)}$ asymptotic
      values. These are obtained by integrating the results on the
      left and subtracting the collinear endpoint contribution
      ($D_\text{end}$ in
      table~\ref{table:multiplicity-coefficients-ee}). The results are
      again split in flavour
      channels.}\label{fig:top-extraction-final}
  \end{subfigure}
  \caption{Illustration of the extraction of the
    $D_{\text{hme},R}^{(bc)}$ coefficients for our inclusive jet
    sample.}\label{fig:top-extraction}
\end{figure}

In this appendix we illustrate how we extract the coefficients
$D_{\text{hme},R}^{(bc)}$ from fixed-order Monte Carlo
simulations. These coefficients are part of the \NNDL resummation
arising from hard matrix-element corrections at the top of the Lund
plane, cf.\ section~\ref{sec:nndl-hard-me}.
We take the example of the inclusive jet sample studied in
section~\ref{sec:pheno} as a reference. An analogous procedure, not
shown here, has been performed for the $Z$+jet sample.

The first step is to run a Monte Carlo simulation for the real part of
the NLO corrections, i.e.\ Born-level $2\to 3$ events in \texttt{NLOJet++}
in the present case.
We build the Lund multiplicity at $\mathcal{O}(\as)$ from these events
split in flavour channels according to the procedure described at the
end of section~\ref{sec:nndl-hard-me}.
The resulting distributions, differential in $\ln k_t$, are shown in
figure~\ref{fig:top-extraction-initial} together with the analytic
\NDL expectations (dashed curves).
The fact that the two converge at large $\ln(p_\perp R/k_t)$ validates
our \NDL calculation.

Next, we subtract the \NDL contribution from the Monte Carlo
distribution and build the cumulative distribution of the remainder.
This is expected to asymptote to the total $\order{\as}$ \NNDL
contribution.
To isolate the piece originating from the top of the Lund plane, we
further subtract the collinear endpoint correction,
$D_\text{end}^{a\to bc}$
(cf.~table~\ref{table:multiplicity-coefficients-ee}).
The resulting (cumulative) distributions are plotted in
figure~\ref{fig:top-extraction-final}. The results have been
normalised by $\frac{\as}{2\pi}$ so as to directly correspond to
$D_{\text{hme},R}^{(bc)}$.  We indeed see a convergence to a constant
value which we extract for each flavour channel by taking the value
at $\ln(p_\perp R/k_t)=7$. The first uncertainty quoted in the legend
of the plot corresponds to the Monte Carlo statistical uncertainty,
and the second number is an estimate of the systematic uncertainties
obtained by extracting $D_{\text{hme},R}^{(bc)}$ at $\ln(p_\perp R/k_t)=6.8$
and 7.2 and taking their difference.
It is delicate to comment on the relative size of the various flavour
channels as only the full sum in Eq.~(\ref{eq:nndl-hme-3}) is
independent of the prescription used to assign flavour.

\bibliographystyle{JHEP}
\bibliography{multiplicity}
\end{document}